%% file: main.tex
\documentclass{fundam}

\usepackage{lineno}
\usepackage{etoolbox}

\usepackage[T1]{fontenc}
\usepackage[utf8]{inputenc}
\usepackage{url}

\usepackage{amssymb,amsmath}
\usepackage{enumerate}
\usepackage{ifthen}
\usepackage[ruled,vlined,linesnumbered]{algorithm2e}
\usepackage{tikz}    
\usepackage{subcaption}
\usepackage{enumitem}
\usepackage{mymacros}

\setlist{nosep}
\setlist[enumerate]{label=\alph*)}

\usetikzlibrary{decorations.pathmorphing}

\begin{document}

\title{Well-Formed Free-Choice Petri Nets Revisited}

\author{%
Petr Jan\v{c}ar\corresponding\\
Department of Computer Science, Faculty of Science, Palack\'y University,\\
Olomouc, Czechia (petr.jancar@upol.cz)
\and Eike Best\\
Department of Computer Science, Carl von Ossietzky Universit\"at Oldenburg,\\
D-26111 Oldenburg, Germany (eike.best@informatik.uni-oldenburg.de)
\and Raymond Devillers\\
D\'epartement d'Informatique, Universit\'e Libre de Bruxelles,\\
B-1050 Brussels, Belgium (raymond.devillers@ulb.be)
\and Mat\v{e}j O\v{s}\v{t}\'{a}dal\thanks{This author was supported by
grant \texttt{No.\,IGA\_PrF\_2026\_016} of Palack\'y University Olomouc.}\\
Department of Computer Science, Faculty of Science, Palack\'y University,\\
Olomouc, Czechia (matej.ostadal01@upol.cz)
}

\runninghead{Jan\v{c}ar, Best, Devillers,
O\v{s}\v{t}\'{a}dal}{Well-Formed Free-Choice Petri Nets Revisited}

\maketitle

\begin{abstract}
	The theory of free-choice Petri nets is an established field,
	initiated in the 1970s by F. Commoner and M. Hack.
	We revisit well-formed free-choice nets (those admitting markings that are both live and bounded) and provide a new characterisation by introducing semi-T-components.
	This notion is dual to that of semi-S-components, which in turn correspond to the well-known minimal siphons.
	By highlighting the symmetry between these dual concepts, we derive the classical coverability theorems for T- and S-components,
	as well as the duality theorem---stating that a free-choice
	net is well-formed if and only if its reverse-dual is also
	well-formed---using
	highly symmetric arguments.
\end{abstract}


\section{Introduction}
Petri nets are a well-known model of parallel and concurrent processes
(see, e.g., the recent monograph~\cite{DBLP:books/sp/BestD24} and the
references therein). Free-choice Petri nets constitute a
well-researched subclass in which behavioural properties, such as liveness and boundedness, are tightly linked to structural constraints, such as coverability by S- and T-components.

The analysis of free-choice Petri nets has a long history, pioneered
in 1972 by F.~Commoner and M.~Hack~\cite{commoner-1972,hack72}.
Subsequently, other authors contributed to the field; most notably,
J.~Esparza, together with M.~Silva and J.~Desel, published a series of
seminal
papers~\cite{DBLP:conf/apn/EsparzaS89a,DBLP:conf/apn/EsparzaS89,
DBLP:conf/concur/Esparza90,DBLP:conf/apn/EsparzaS90} and a
monograph~\cite{de95}. 
Later, 
several authors noted (see, e.g.,
\cite{DBLP:journals/jcss/GaujalHM03,DBLP:journals/fuin/Wehler10,DBLP:journals/fuin/Aalst21}) that certain classical arguments warrant further revision, extension, and application.

Free-choice nets form the analysable core of workflow
nets---the de facto standard model in business process management and
process mining.
The central property of workflow nets, soundness,
amounts to liveness and boundedness of the ``short-circuited''
net~\cite{DBLP:journals/fac/AalstHHSVVW11}, meaning that
well-formedness,
the concept which is central to the present paper, serves
precisely as the structural core of soundness.

Free-choice and related structural restrictions also play a role in hardware and
control applications: in the synthesis of asynchronous circuits~\cite{CortadellaKKLY02} and the decomposition of concurrent controllers into
state-machine components (i.e., covers by S-components) for
verification~\cite{DBLP:journals/amcs/WojnakowskiWBP21}.

Beyond practical modeling, the theory of free-choice nets remains an active research area.
Recent lines of work
include the study of deterministic negotiations---a model equivalent to
free-choice workflow nets---for which soundness is NL-complete~\cite{DBLP:journals/lmcs/EsparzaKMW18};
new results on perpetual free-choice nets~\cite{DBLP:conf/apn/Aalst18};
and fresh complexity results concerning reachability and alignments in sound free-choice workflow nets~\cite{DBLP:conf/apn/PrinzSA25, DBLP:conf/apn/SchwanenPA25}.

Before detailing our contribution, we recall that several categories of arguments may motivate the significance of a paper:
\begin{enumerate}[label=\arabic*)]
    \item New results are developed with interesting consequences.
    \item New algorithms are devised to check important properties efficiently and transparently, or to construct relevant objects.
    \item New proofs are presented for established results. Although often under-appreciated, we believe such contributions are of substantial interest if they shed new light on the subject, or if the proofs are more concise, accessible, elegant, or exhibit surprising characteristics (such as inherent symmetry).
\end{enumerate}
The present work is primarily concerned with item~3. We focus on the
characterisation of well-formed free-choice nets (that is, those
admitting a live and bounded marking) through their structural
subnets, specifically \emph{semi-T-components} and their duals,
\emph{semi-S-components}. 
These structures are introduced in Section~\ref{sec:main} as approximations of the better-known concepts of T-components and
S-components. A~semi-S-component is akin to the place-based notion of
\emph{minimal siphons} (cf.~\cite{de95}).
Symmetrically, a~semi-T-component is akin to a transition-based notion which, however,
does not appear to have been widely adopted in the literature.

We show that the notions of semi-T-components and
semi-S-components lend themselves to a variety of duality arguments.
In particular, it turns out that a strongly connected (not necessarily well-formed)
free-choice net is covered by both semi-T-components (Proposition~\ref{prop:semiTcover})
and semi-S-components (Proposition~\ref{prop:semiScover}); this is demonstrated using purely structural (i.e., graph-theoretical) arguments.
If strong connectedness is strengthened to well-formedness, these two
results can be further refined by omitting the ``semi'' prefix in both
cases. This retrieves, via Theorem~\ref{thm:wfChar}, the classical S-coverability and T-coverability
results~\cite{hack72,de95}, along with the classical duality result as
a direct consequence (originally established by Hack, this states that
a free-choice net is well-formed if and only if its reverse-dual net is well-formed).
We note that the introduction of semi-T-components and their application in Theorem~\ref{thm:wfChar} may also be regarded as a contribution of the first type (as per the categories outlined above), yielding new results with interesting consequences.

Several polynomial-time algorithms exist for
checking the well-formedness of free-choice nets.
Some of 
them are based on linear algebraic methods, relying on 
the Rank
Theorem~\cite{DBLP:conf/apn/EsparzaS89a,DBLP:conf/apn/Desel92,DBLP:conf/apn/TeruelS94}.
These include an algorithm which, to the best of our knowledge, offers the
lowest complexity~\cite{DBLP:conf/apn/KemperB92,kemper_cover}.
Other algorithms, such as the ones proposed
in~\cite{DBLP:conf/apn/BarkaouiM92, DBLP:conf/apn/BarkaouiCD95} are
based on graph theory.
In fact, using our characterisation of well-formedness via
semi-T-components, we present
in Appendix~\ref{sec:alg} a polynomial-time
algorithm which
uses a subprocedure to decide structural liveness that is dual
to the algorithm of~\cite{DBLP:conf/apn/BarkaouiCD95}.
In our presentation, we prioritise a
concise and transparent demonstration of its correctness over algorithmic optimization.
Our algorithm may therefore be regarded as a modest contribution of
the second type (as per the categories outlined above),
although we cannot claim any fundamental novelty here.

\emph{Organization of the paper.} Classical definitions regarding Petri nets are recalled in Section~\ref{sec:prelim}, emphasizing graph-theoretical aspects while introducing key behavioral concepts, such as boundedness and liveness, alongside the structural concept of well-formedness. 

Section~\ref{wfsc.sct} provides a self-contained proof of the established fact that well-formedness implies strong connectedness (for potentially pairwise unconnected net components). This section is included because the underlying proof methods are subsequently employed in the main body of the paper.

Section~\ref{sec:main} forms the central part of this paper; it develops the characterisation of well-formed free-choice nets, culminating in Theorem~\ref{thm:wfChar}. 
Concluding remarks are presented in Section~\ref{sec:remarks}, referencing Appendix~\ref{sec:alg} for the aforementioned well-formedness algorithm.

Appendix~\ref{commoner.app} is derived
from~\cite{DBLP:journals/corr/abs-2401-12067} and demonstrates that
the proof methods of the main part can be employed to provide an
independent proof of the well-known theorem by F.~Commoner
characterising the liveness of free-choice nets. We note that in more
standard developments~\cite{de95,DBLP:books/sp/BestD24}, Commoner's
Theorem is typically used as a stepping stone to prove coverability
theorems; our approach isolates the part that is really used, via
Proposition~\ref{prop:noProperSwf}. Moreover, the proof presented here
appears to be somewhat less involved than those in standard texts, as it avoids the technicalities of an explicit transition-allocation.

This paper arises as a revision and extension
of the conference paper~\cite{DBLP:conf/apn/BestDJ25}. In particular,
the presentation has been simplified so that the duality theorem follows
immediately, and the algorithm has been integrated.

\section{Preliminaries}\label{sec:prelim}

We denote the set of integers by
$\setZ$, and 
the set of nonnegative integers $\{0,1,2,\dots\}$
by $\setN$.
For any $i,j\in\setN$, we define the interval
$[i,j]=\{i,i{+}1,\dots,j\}$, which is empty if $i>j$.
For a function $f\colon A\to B$ and $A'\subseteq A$, the symbol $f\restr{A'}$
denotes the restriction of $f$ to the subdomain $A'$.

Given a~set $T$, let $T^*$ and $T^\omega$ denote the
set of finite sequences and the set of infinite sequences of elements
from $T$, respectively (where $\omega$ refers to the least infinite
ordinal); the symbol $\varepsilon$ denotes the empty sequence.
For a sequence $\sigma\in T^*\cup T^\omega$ and a subset $T'\subseteq T$, 
the restriction of $\sigma$ to $T'$, denoted by $\sigma\restr{T'}$,
is obtained from $\sigma$ by removing all elements that are not in
$T'$.

\paragraph{Notions from graph theory.}
Let $G=(V,E)$ be a~(directed) \emph{graph}, where $V$ is the set of
\emph{nodes}
(or vertices) and $E \subseteq V \times V$ is the set of
\emph{arcs}. 
For $V'\subseteq V$, $G[V']$ denotes the \emph{subgraph of} $G$
\emph{induced by} $V'$, that is, $G[V']=(V',E\cap (V'\times V'))$. 

If $V=V_1\sqcup V_2$ (disjoint union) such that $G[V_1]$ and $G[V_2]$ have
no arcs, then $G$ is a \emph{bipartite graph}
(in which $E\subseteq (V_1\times V_2)\cup (V_2\times
V_1)$).

A~\emph{path} of \emph{length} $k$ \emph{from} $x$ \emph{to} $y$ is a
sequence of nodes $x_0, x_1, \dots, x_k$ where $x=x_0$, $x_k=y$, and
$(x_i, x_{i+1}) \in E$ for all $i\in[0,k{-}1]$. We also say
that the \emph{arc} $(x_i,x_{i+1})$ \emph{belongs to} the
\emph{path} for
$i\in[0,k{-}1]$.
A \emph{path} $x_0, x_1, \dots, x_k$ is
\emph{simple} if its nodes are distinct ($x_i\neq x_j$ for
$i\neq j$). 
A~\emph{cycle} is a path  $x_0, x_1, \dots, x_k$ where $k>0$ and
$x_0=x_k$.
A \emph{cycle}  $x_0, x_1, \dots, x_k$ is \emph{simple} if the path $x_0, x_1,
\dots, x_{k-1}$ is simple.

The \emph{distance from} a \emph{node} $x$ \emph{to} a \emph{node} $y$
is the length of a shortest
path from $x$ to $y$, if such a~path exists; otherwise,
it is infinite.
The \emph{distance from} a \emph{node} $x\in V$ \emph{to} a \emph{set} $V'\subseteq
V$ (or \emph{to} a \emph{subgraph} $G[V']$) 
is the minimum distance from $x$ to any node in $V'$; the
distance from $x$ to the empty set $\emptyset$ is infinite.

A graph $G=(V,E)$ is \emph{strongly connected} if
for every pair of nodes $x,y\in V$ there exists a path from $x$ to $y$.
A~nonempty subgraph $G[V']$ is 
a~\emph{strongly connected component} (\emph{SCC} for short) \emph{of}
$G$ if $G[V']$ is strongly connected and is maximal with respect to
inclusion; that is,
there is no strongly connected subgraph $G[V'']$ such that
$V'$ is a proper subset of $V''$.

An~SCC $X=G[V']$ of $G$ is a~\emph{top SCC} \emph{of} $G$ if it has no
incoming arcs in $G$; that is, there are no arcs  $(x, y) \in E$
such that
$x \notin V'$ and $y \in V'$.
Symmetrically, $X$ is a~\emph{bottom SCC} if it has no
outgoing arcs in $G$; that is, there are no arcs $(x, y) \in E$ such that
$x\in V'$ and $y\notin V'$.

We recall that there are well-known algorithms (such as Tarjan's
algorithm~\cite{Tarjan72}) 
that, given a finite graph $G=(V,E)$, construct all SCCs of $G$ in 
$O(|V|{+}|E|)$ time.

\DEF{def:petrinets}{Place/transition Petri net, preset, postset, notation $N=(S_N,T_N,F_N)$}
	A~\emph{Petri net}, or simply a \emph{net}, is a~triple $N=(S,T,F)$ where
	$S$ and $T$ are finite disjoint sets of \emph{places} and \emph{transitions}, respectively,
	and $F \subseteq (S \times T) \cup (T \times S)$ is the
	\emph{flow relation}.
	For a \emph{node} $u \in S\cup T$, we define
	$\pre{u} = \{v \mid (v,u) \in F\}$ (the \emph{preset of $u$}),
	and 
	$\post{u} = \{v \mid (u,v) \in F\}$ (the \emph{postset of $u$}).
	If $\pre{u} = \post{u} = \emptyset$,
	then $u$ is an \emph{isolated node}.
	For a~subset $U \subseteq S \cup T$,
	we define $\pre{U} = \bigcup_{u \in U} \pre{u}$ and 
	$\post{U} = \bigcup_{u \in U} \post{u}$.
		
  	For any net $N$, we denote its constituent sets by $S_N$, $T_N$, and
  	$F_N$; thus $N = (S_N, T_N, F_N)$.
\ENDDEF

\begin{remark}
	In the literature, the set of places is frequently denoted by $P$. 
	Following the monograph~\cite{de95}, we use $S$ (local states),
	a notation stemming from the German word \emph{Stellen}.
	It should also be noted that this paper deals exclusively with \emph{plain nets}, where
	the arcs in $F$ are of 
	weight $1$.
	In the case of nets with \emph{weighted arcs}, $F$ would instead be
	viewed as a function
	$F\colon (S \times T) \cup (T \times S)\to\setN$.
\end{remark}

Since a net $N=(S,T,F)$ defines a directed bipartite graph $(S\cup T, F)$,
we can apply the graph-theoretic notions introduced above to nets. The subgraph
notation $G[V']$ also motivates the following definition of a subnet.

\DEF{def:subnets}{Subnet and the dot notation}
	Given a net $N$ and a set $U\subseteq S_N\cup T_N$, we denote by $N[U]$
	the net $(U\cap S_N, U\cap T_N, F_N\cap (U\times U))$, called
	the \emph{subnet of} $N$ \emph{induced by} $U$.

	If we consider a subnet $X$ of a net $N$, the dot notation 
	($\pre{u}, \pre{U}, \post{u},\post{U}$) refers to the flow relation in
	$N$, not in $X$. (For example,
	$\pre{S_X}$ might contain transitions from $T_N\setminus T_X$.)  
\ENDDEF

We can thus freely use terms such as an \emph{SCC} $X$ \emph{of} a
\emph{net} $N$, which refers to an SCC of the graph $(S_N\cup T_N,
F_N)$ as well as to the induced 
subnet $X=(S_X,T_X,F_X)=N[S_X\cup T_X]$.

\DEF{def:markings}{Marking, enabling, firing, effect $\Delta(\sigma)$, execution $M\gt{\sigma}$, reachability set $\rset{M_0}$}
	A \emph{marking} of a net $N = (S, T, F)$ is a function $M \colon S
	\to \setN$, attaching a number $M(s)$ of \emph{tokens} (the marking of
	$s$) to each place $s \in S$. 
	The symbol $\mathbf{0}$ denotes the \emph{zero marking}, defined as
	$\mathbf{0}(s) = 0$ for all $s \in S$. 
	For any $S' \subseteq S$, we may write $M\restr{S'} = \mathbf{0}$ as a
	shorthand for $M\restr{S'} = \mathbf{0}\restr{S'}$;
	in this case, we also say that $S'$ is \emph{unmarked at} $M$.

	A \emph{transition} $t \in T$ is \emph{enabled at} a marking $M$ if
	$M(s) \geq 1$ for all places $s \in \pre{t}$; this is denoted by $M
	\gt{t}$. The \emph{effect} $\Delta(t) \colon S \to \setZ$ of a transition $t$ is defined as:
	$\Delta(t)(s) = -1$ if $s \in \pre{t} \setminus \post{t}$, 
	$\Delta(t)(s) = +1$ if $s \in \post{t} \setminus \pre{t}$, and 
	$\Delta(t)(s) = 0$ otherwise. 
	An enabled transition $t \in T$ may \emph{fire} at a~marking $M$, leading to a new marking $M' = M + \Delta(t)$; this is denoted by $M \gt{t} M'$.

	We extend this notation to $M \gt{\sigma}$ for firing sequences $\sigma \in T^* \cup T^\omega$, and to $M \gt{\sigma} M'$ for finite sequences $\sigma \in T^*$ (defined inductively by $M \gt{\varepsilon} M$, and $M \gt{\sigma t} M''$ if $M \gt{\sigma} M'$ and $M' \gt{t} M''$).
	Note that 
	\begin{center}
		for $M \gt{\sigma} M'$ we have $M'=M+\Delta(\sigma)$, where 
	$\Delta(t_1t_2\cdots t_m)=\sum_{i=1}^m\Delta(t_i)$.
	\end{center}
	We also refer to $M \gt{\sigma}$ or $M \gt{\sigma} M'$ as an
	\emph{execution} of $N$, which may be finite or infinite.

	For an (initial) marking $M_0$, we define its \emph{reachability set} 
	(the set of all markings \emph{reachable from} $M_0$)
	as
	\[
	\rset{M_0} = \{ M \mid M_0 \gt{\sigma} M \text{ for some } \sigma \in T^* \}.
	\]
\ENDDEF

\begin{figure}[t]
    \centering
	\input{figures/wf_sc.tex}
    \caption{A non-strongly connected net consisting of two SCCs separated by the dashed arcs.}
    \label{fig:wf_sc}
\end{figure}
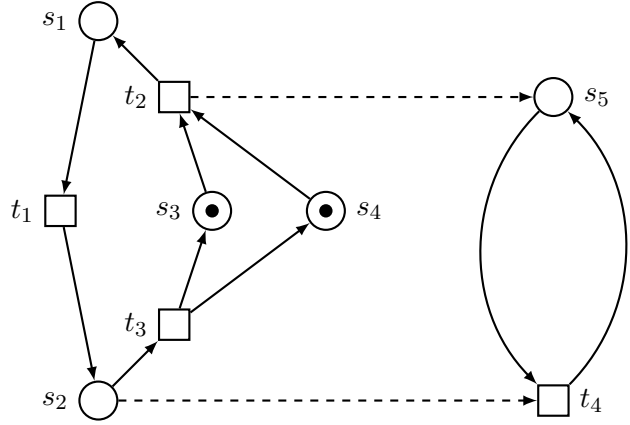

\DEF{def:wellformed}{Bounded marking, live marking, well-formed net}
	Given a net $N=(S,T,F)$,
	a \emph{marking} $M_0$ is \emph{bounded} if  its reachability set
	$\rset{M_0}$ is finite; that is,  
	if there exists a~bound $b\in\setN$ such that $M(s)\leq b$ for all 
	$M\in\rset{M_0}$ and all places $s\in S$.
	In this case we also say that the \emph{system} $(N,M_0)$ is
	\emph{bounded}.

	A \emph{transition} $t\in T$ is \emph{live} at a marking $M$ if 
	for every marking $M' \in \rset{M}$, there exists a firing sequence
	$\sigma \in T^*$ such that $M' \gt{\sigma t}$.
	A \emph{marking} $M_0$ is \emph{live} if every transition $t \in T$ is
	live at $M_0$. In this case we also say that the \emph{system} $(N,M_0)$ is
	\emph{live}.

	A \emph{net} $N$ is \emph{well-formed} if there exists a marking $M_0$
	that is live and bounded (that is, the system $(N, M_0)$ is both live and bounded).
\ENDDEF

\DEF{def:deadtrans}{Dead transition, DL-marking}
	Given a net $N=(S,T,F)$, a \emph{transition} $t \in T$ is \emph{dead} at a marking $M$ if there is no firing sequence $\sigma \in T^*$ such that $M \gt{\sigma t}$.
	A marking $M$ is called a \emph{DL-marking} if every transition $t \in
	T$ is either dead or live at $M$, and at least one transition is dead.
	A particular DL-marking is a \emph{dead marking} (or a
	\emph{deadlock});
	that is, a marking $M$ at which all transitions are dead, provided
	that $T\neq\emptyset$.
\ENDDEF

Note that, by definition, a transition $t \in T$ is not live at $M$ if and only if $t$ is dead at some $M' \in \rset{M}$.
We recall some further standard facts:

\PROP{prop:dlmarking}{Liveness and DL-markings}
	Given a net $N=(S,T,F)$ with $T \neq \emptyset$:
	\begin{enumerate}[label=\arabic*)]
    	\item If a marking $M_0$ is live, then every $M \in \rset{M_0}$ is also live. Consequently, there exists an infinite execution $M_0 \gt{\sigma}$ in which every transition $t \in T$ fires infinitely often.
    	\item A marking $M_0$ is live if and only if no DL-marking is reachable from $M_0$.
	\end{enumerate}
\ENDPROP

\PF
	1. This follows immediately from the definition of liveness: if $M_0$ is live, 
	then every transition is fireable from any reachable marking.
	The existence of an infinite execution from $M_0$ in which every
	transition fires infinitely often is a direct consequence of this property.

	2. ($\Rightarrow$) If $M_0$ is live, every $t \in T$ is live
	at every $M \in \rset{M_0}$. Since a DL-marking requires at
	least one transition to be dead, no such marking can be
	reachable from $M_0$.

	($\Leftarrow$) Let $L(M)$ and $D(M)$ denote the sets of transitions
	that are live and dead at $M$, respectively. Note that for any $M' \in
	\rset{M}$, we have $L(M) \subseteq L(M')$ and $D(M) \subseteq D(M')$.
	Moreover, for any transition $t \in T \setminus (L(M) \cup D(M))$,
	there exists $M' \in \rset{M}$ such that $t \in D(M')$.
	This readily implies that if $M_0$ is not live (that is, there exists
	$t \in T \setminus L(M_0)$), then a DL-marking $M$ is necessarily
	reachable from $M_0$.
\ENDPF

\begin{example}
	Figure~\ref{fig:wf_sc} shows a net $N$ with five places 
	and four transitions. 
	As is standard, places are represented by circles and transitions by boxes.
	The graph of $N$ has two SCCs (a top
	SCC on the left and a bottom SCC on the right).
	The figure also depicts a marking $M_0$ of $N$, where each place contains a number of black tokens; this marking can be represented as the vector $(0,0,1,1,0)$.

	An execution of $N$ is, for instance:
	\[
	(0,0,1,1,0) \gt{t_2} (1,0,0,0,1) \gt{t_1t_3} (0,0,1,1,1)
	\gt{t_2t_1t_3} (0,0,1,1,2) \gt{t_2t_1t_4} (0,0,0,0,3).
	\]
	The execution demonstrates that $(N, M_0)$ is neither bounded
	(consider $\sigma=(t_2t_1t_3)^\omega$)
	nor live; the marking $(0,0,0,0,3)$ is a DL-marking, at which
	even all transitions are dead.

	We may also note that each SCC constitutes a well-formed subnet.
	However, if the arc $(t_2, s_2)$ or $(s_2, t_2)$ were added, the top SCC would no longer be well-formed.
\end{example}

\subsection{Well-formedness implies strong connectivity} 
\label{wfsc.sct} 

Theorem~\ref{th:wfImpSc} provides an alternative proof of a well-known
result (cf.~Theorem 2.25 in~\cite{de95}). This proof is based
on Proposition~\ref{prop:omitBotSCC}, which deals with bottom SCCs;
the proposition is also utilised in Section~\ref{sec:main}. We note
that while our current context is limited to plain nets, the proof
remains valid for nets with weighted arcs as well.

\PROP{prop:omitBotSCC}{Transitions from a bottom SCC may safely be omitted}
	Let $B$ be a bottom SCC of a net $N$.
	For any execution $M \gt{\sigma}$\ of $N$,
	whether finite or infinite, there exists an execution $M\gt{\sigma'}$, where
	$\sigma'$ is obtained from $\sigma$ by omitting all transitions
	from $T_B$; that is, $\sigma'=\sigma\restr{T'}$ for $T'=T_N\setminus T_B$.
\ENDPROP
\PF
	Let $B$ be a bottom SCC of $N$; we define $S'=S_N\setminus S_B$
	and $T'=T_N\setminus T_B$. 
	Since $B$ is a bottom SCC, for each  $t\in T_B$ we have
	$\post{t}\subseteq S_B$. Consequently,
	$\Delta(w)\restr{S'}\leq \mathbf{0}$ for all transition sequences
	$w\in (T_B)^*$; that is, $\Delta(w)(s)\leq 0$ for all $s\in S'$.
	On the other hand,
	for each $t'\in T'$, we have $\pre{t'}\subseteq S'$.

	Suppose, for the sake of contradiction,
	that $M\gt{\sigma}$ is an execution but $M\gt{\sigma'}$ is not; that
	is, $\sigma'=\sigma\restr{T'}$
	is not enabled at $M$. Then $\sigma$ must have a finite
	prefix $w_0t'_1w_1t'_2\cdots w_{m-1}t'_mw_mt'_{m+1}$
	with $w_i\in(T_B)^*$ for all $i\in[0,m]$ and $t'_j\in T'$ for all
	$j\in[1,m{+}1]$, such that
	\begin{center}
		$M\gt{w_0t'_1w_1t'_2\cdots w_{m-1}t'_mw_m}M'\gt{t'_{m+1}}$, 
	\ $M\gt{t'_1t'_2\cdots t'_m}M''$, and
	\ $t'_{m+1}$ is disabled at $M''$.
	\end{center}
	Since $M''=M'-\Delta(w_0w_1\cdots w_m)$ and 
	$\Delta(w_0w_1\cdots w_m)\restr{S'}\leq\mathbf{0}$, it follows that  $M''\restr{S'}\geq M'\restr{S'}$.
	Since $M'\gt{t'_{m+1}}$ and $\pre{t'_{m+1}}\subseteq S'$, 
	the transition $t'_{m+1}$ must also be enabled at $M''$, yielding a contradiction.
\ENDPF

\THM{th:wfImpSc}{Well-formedness implies strong connectivity}
	In a well-formed net,
	every bottom SCC is also a~top SCC.
	Consequently, every well-formed net is either strongly connected or consists of a collection of pairwise unconnected, strongly connected well-formed components.
\ENDTHM
\PF
	Let $B$ be a bottom SCC of a net $N$ that is not a top SCC; 
	hence, $F_N$ contains an arc $(x,y)$ such that
	$x\notin S_B\cup T_B$ and $y\in S_B\cup T_B$.
	We aim to show that $N$
	is not well-formed. 
	Specifically, we demonstrate that the existence of an
	infinite execution $M\gt{\sigma}$ of $N$ satisfying  
	$T^\sigma_\infty=T_N$  (where $T^\sigma_\infty$ denotes the set of transitions
	occurring infinitely often in $\sigma$) 
	implies that the marking $M$ is unbounded.
	Since such an infinite execution must exist
	for every live marking $M$, it follows that $N$ cannot possess a marking that is both live and bounded.

	We consider an
	infinite execution $M\gt{\sigma}$ of $N$ satisfying
	$T^\sigma_\infty=T_N$, and 
	show that the reduced execution $M\gt{\sigma'}$ with
	$\sigma'=\sigma\restr{T_N\setminus T_B}$ (which exists by
	Proposition~\ref{prop:omitBotSCC}) demonstrates that $M$ is unbounded.
	We fix an arc 
	$(x,y)$ such that $x\notin S_B\cup T_B$ and $y\in S_B\cup T_B$,
	and distinguish two cases based on the type of the arc $(x,y)$
	(the two types are demonstrated by the arcs $(t_2,s_5)$ and $(s_2,t_4)$ in Figure~\ref{fig:wf_sc}):

	Case 1: $x$ is a transition, and $y$ is a place
	($x\in T_N\setminus T_B$, and $y\in S_B$).
	In the reduced execution $M\gt{\sigma'}$, the marking of $y$ is infinitely
	often increased
	(whenever $x$ fires). 
	Since $B$ is a bottom SCC, we have $\post{y}\subseteq
	T_B$. Because no transitions from $T_B$ occur in $\sigma'$, the
	marking of $y$ is never decreased in the execution $M\gt{\sigma'}$, and thus grows above any bound.

	Case 2: 
	$x$ is a place, and $y$ is a transition 
	($x\in S_N\setminus S_B$, and $y\in T_B$).
	In the original execution $M\gt{\sigma}$,
	the marking of $x$ is infinitely often decreased by $y\in T_B$ 
	and never increased by any transition from
	$T_B$, since $\pre{x}\cap T_B=\emptyset$.  Hence, the reduced execution $M\gt{\sigma'}$ omits infinitely
	many decreases of the marking of $x$ while keeping all increases.
	This causes the marking of $x$ to grow above any bound in
	$M\gt{\sigma'}$.
\ENDPF

\section{Well-formed free-choice nets}
\label{sec:main}

The main results of this section are captured by Theorem~\ref{thm:wfChar}:
a free-choice net is well-formed if and only if it is covered by
T-components and none of its semi-T-components is proper (that is, all of them are
T-components)---and if and only
if the same holds with S in place of T. The classical coverability theorems
are the forward implications, and the duality theorem is a direct consequence.

\subsection{Free-choice nets}

We recall the standard definition of free-choice nets. These nets are composed of interconnected subnets called clusters. See Figure~\ref{fig:five_clusters_and_bott_scc} for an example.

\DEF{def:fc}{Free-choice net, cluster} 
	A net $N$ is a~\emph{free-choice net} if
	$\pre{t_1}\cap\pre{t_2}\neq\emptyset$ implies $\pre{t_1} =
	\pre{t_2}$,
	for all $t_1, t_2\in T_N$.

	A \emph{cluster} of $N$ is a subnet $C = N[S_C \cup T_C]$ such that
	$\post{s} = T_C$ for every $s \in S_C$ and $\pre{t} = S_C$ for every $t \in T_C$, where $S_C \cup T_C$ is an inclusion-maximal set satisfying this property. 

	By $\clusters_{N}$ we denote the set of clusters of a free-choice net
	$N$. 
	For any node $x\in S_N\cup T_N$, let $C(x)\in\clusters_N$ denote
	the cluster containing $x$.  
\ENDDEF

The inclusion-maximality ensures that all places $s$ with
$\post{s}=\emptyset$ belong to a single cluster; analogously, the same
applies to all
transitions $t$ with $\pre{t}=\emptyset$. Note the following standard
(and easily verifiable) facts, where item 1 confirms that 
the cluster $C(x)$ is well-defined.

\PROP{prop:subFree}{Cluster partition; subnets
	are free-choice; either all $t\in T_C$ enabled, or none}
	Given a free-choice net $N$:
	\begin{enumerate}[label=\arabic*)]
		\item
			The set of nodes $S_N\cup T_N$ is partitioned into
			clusters; that is, each $x\in S_N\cup T_N$ belongs to
			exactly one cluster $C(x)\in\clusters_N$.
		\item		
			Any subnet of $N$ is itself a free-choice net.
		\item
			For any marking $M$ and any cluster $C\in\clusters_N$, either all
			transitions in $T_C$ are enabled at $M$, or none of
			them are.	
	\end{enumerate}
\ENDPROP

\subsection{Characterisation of well-formed free-choice nets by semi-components}

To understand \emph{well-formed} free-choice nets, it suffices to
consider
strongly connected free-choice nets (by Theorem~\ref{th:wfImpSc})
that contain at least one place and
at least one transition (as other cases are trivial).
In such nets,
each cluster necessarily contains at least one transition
and at least one place.
In light of the all-or-none property (Proposition~\ref{prop:subFree}(3)), it proves useful to examine subnets formed by selecting exactly one transition from each cluster and removing all transitions not chosen.

\DEF{def:alloc}{Allocation $\alpha$, induced subnet
    $N_\alpha$, directed allocation}
	Given a~free-choice net $N$ where each cluster contains at least one
	transition,
	an \emph{allocation} is a function
	$\alpha: \clusters_N \rightarrow T_N$ such that $\alpha(C)\in T_C$ for
	each cluster $C$.

	Given an allocation $\alpha$, we denote by $N_\alpha$ the subnet
	induced by the set $S_N$ of all places of $N$ and the set
	$T_{N_\alpha}=\{\alpha(C)\mid C\in\clusters_N\}$ of all chosen
	transitions.

	An \emph{allocation} $\alpha$ is \emph{directed to} a \emph{node} $u \in S_N
	\cup T_N$ if from every node of $N_\alpha$ there is a path to $u$ in
	$N_\alpha$ (which implies that $u\in S_N\cup T_{N_\alpha}$). It is a \emph{directed allocation} if it is directed to
	some node (that is, if $N_\alpha$ has exactly one bottom SCC).
\ENDDEF

\begin{remark}\label{rem:bottomYtopX}
	By the definition above, an allocation is a
	\emph{transition-allocation} $\alpha$ which selects exactly one transition from each cluster.
	We will be particularly concerned with the \emph{bottom SCCs} $Y$ of the nets
	$N_\alpha$. 
	As a motivating observation,  anticipating the duality to be discussed
	later,
	we note that it will subsequently be
	shown (see Proposition~\ref{prop:MDLtopSCC})
	that for every DL-marking $M$, there exists  a \emph{top SCC}
	$X$ of $N_\beta$ such that
	$M\restr{S_X}=\mathbf{0}$, where $\beta$ is a \emph{place-allocation}
	selecting
	one place in each cluster.
\end{remark}

We will observe that any strongly connected free-choice net $N$ 
(with $T_N\neq\emptyset$)
is covered
by the bottom SCCs of its subnets $N_\alpha$ induced by directed
allocations $\alpha$.
This motivates the following definition of
\emph{semi-T-components} of $N$, which correspond to the bottom SCCs
of the subnets $N_\alpha$.
We introduce this notion for general nets as an extension of
the standard notion
of \emph{T-components}. It will turn out that for well-formed
free-choice nets, every semi-T-component is in fact a T-component. 

\DEF{def:semiTcomp}{Semi-T-component of net $N$}
	A~subnet $Y$ of a net $N$ is a~\emph{semi-T-component of} $N$ 
	if $T_Y \neq \emptyset$ and the following conditions hold
	(where the pre- and post-functions ``$\pre{}$'' 
	refer to $N$, by our convention):
	\begin{enumerate}
		\item
			$Y$ is strongly connected (which implies
			$S_Y\subseteq \pre{T_Y}\cap\post{T_Y}$);
		\item
			$|\post{s} \cap T_Y|=1$ for each $s\in
			S_Y$ (note that $|\pre{s} \cap T_Y|\geq 1$
			follows from strong connectivity); 
		\item
			$\post{T_Y}\subseteq S_Y$ (hence  
			$\post{T_Y}=S_Y\subseteq\pre{T_Y}$).
	\end{enumerate}
\ENDDEF

\begin{figure}
    \centering
	\vspace{-5em}
    \begin{subfigure}{\linewidth}
        \centering
        \input{figures/five_clusters.tex}
        \label{fig:five_clusters}
    \end{subfigure}

    \vspace{-8em}

    \begin{subfigure}{\linewidth}
        \centering
        \input{figures/bott_scc_alloc.tex}
        \label{fig:bott_scc_alloc}
    \end{subfigure}
	\vspace{-5em}
    \caption{The upper panel shows a~strongly connected free-choice
	net $N$ with five clusters, highlighting a proper semi-T-component of
	Type II but not Type I.
	The lower panel illustrates the 
	net $N_{\alpha}$, where the set of $\alpha$-allocated
	transitions is
	$\{t_{11}, t_{21}, t_{31}, t_{42}, t_{51}\}$.
	Its bottom SCC (depicted in bold) is a~proper semi-T-component of Type I (
	with $(t_{21}, s_{11})$ viewed as an ``excessive arc'')
	and also of Type II (due to the ``inbound arc'' $(s_{32}, t_{31})$).}
	\label{fig:five_clusters_and_bott_scc}
\end{figure}
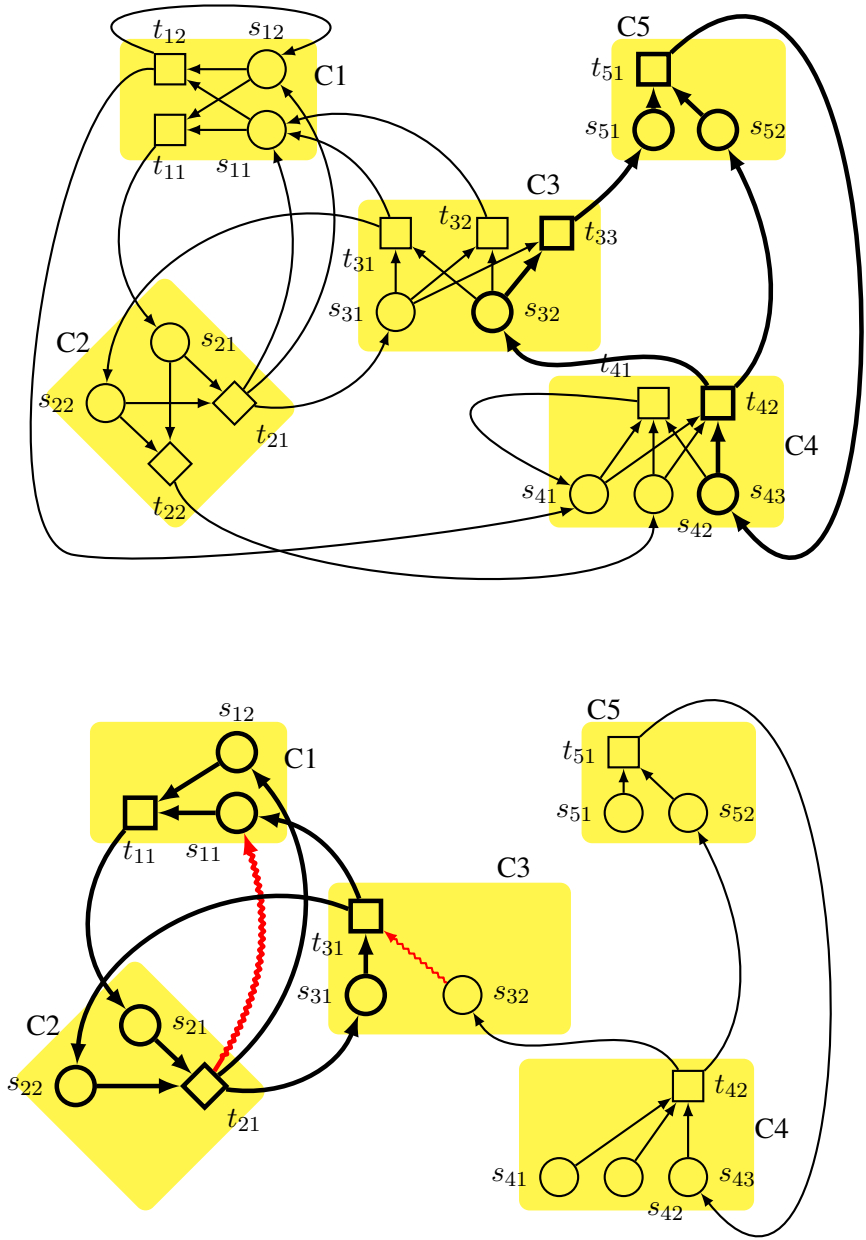

\begin{example}
	The upper panel of Figure~\ref{fig:five_clusters_and_bott_scc} shows a
	semi-T-component in bold. The lower panel highlights another
	semi-T-component in bold (the wavy arcs will be discussed
	later); in this case, it is explicitly depicted as the bottom SCC of
	$N_\alpha$ for a directed allocation $\alpha$.
\end{example}	

\PROP{prop:botSemiT}{Bottom SCCs of $N_\alpha$ coincide with semi-T-components of $N$}
	Given a~free-choice net $N$ where each cluster contains at least one
	transition, a subnet $Y$ is a semi-T-component of $N$ if and only if
	$Y$ is a bottom SCC of $N_\alpha$ for some allocation $\alpha$.
	If $N$ is, moreover, strongly connected, then any semi-T-component
	is the bottom SCC of $N_\alpha$ for some \emph{directed} allocation $\alpha$.
\ENDPROP
\PF
	We start with the ``if'' direction ($\Leftarrow$) of the claimed
	equivalence. Let $B$ be a bottom
	SCC of $N_\alpha$ for some allocation $\alpha$; by definition, 
	$S_B\cup T_B\neq \emptyset$ and
	the subnet
	$B$ is strongly connected.
	Since $\alpha$ is an allocation, 
	each cluster of $N_\alpha$
	contains precisely one
	transition.
	For any place $s \in S_N$, let $t_s$ denote its unique successor in $N_\alpha$, that is, the single transition satisfying $(s,t_s) \in F_{N_\alpha}$. 

	Since $B$ is a \emph{bottom} SCC, no arcs in $N_\alpha$ leave $B$.
	Thus, for each $s \in S_B$, we have $t_s \in T_B$, which implies
	$|\post{s} \cap T_B| = 1$. Together with $S_B\cup T_B\neq \emptyset$,
	this also ensures $T_B \neq \emptyset$ (even if $S_B=\emptyset$).
	Finally, since there are no arcs $(t,s)\in F_N$ such that $t\in T_B$ and $s\notin
	S_B$, we have
	$\post{T_B} \subseteq S_B$. We have established that $B$ is a
	semi-T-component.

	For the ``only-if'' direction ($\Rightarrow$), let $Y$ be a semi-T-component of
	$N$. Since $|\post{s}\cap T_Y|=1$ for each $s\in S_Y$, every cluster of
	$N$ contains at most one transition from $T_Y$.
	We define an allocation $\alpha$ such that $\alpha(C) = t$ whenever $T_C \cap T_Y = \{t\}$, and $\alpha(C)$ is chosen arbitrarily if $T_C \cap T_Y = \emptyset$. Then $Y$ is a strongly connected subnet of $N_\alpha$, and since no arc leaves $Y$ in $N_\alpha$ (as $\post{T_Y} \subseteq S_Y$), $Y$ is a bottom SCC of $N_\alpha$.

	Moreover, if $N$ is strongly connected and we choose $\alpha$ so that
	$\alpha(C)$ is a transition from $T_C$ with the shortest
	distance to $Y$ (thus $\alpha(C) = t$ whenever $T_C \cap T_Y =
	\{t\}$), then $\alpha$ is a directed allocation and $Y$ is the bottom
	SCC of $N_\alpha$.
\ENDPF

\PROP{prop:semiTcover}{Strongly connected free-choice net is covered by semi-T-components}
	Every strongly connected free-choice net $N$ is covered by
	its semi-T-components; that is, each $t\in T_N$ belongs to
	$T_Y$ for some
	semi-T-component $Y$ of $N$. 
\ENDPROP
\PF
	Given a strongly connected free-choice net $N$ and $t_0 \in T_N$, we define an allocation $\alpha$ such that for each cluster $C$, $\alpha(C)$ is a transition in $T_C$ with a shortest distance to $t_0$ in the graph $N$. In particular, $\alpha(C(t_0)) = t_0$. 

	By this construction, from every node in $N_\alpha$ there exists a path to $t_0$ in $N_\alpha$. Hence $\alpha$ is a directed allocation, and $t_0$ belongs to the bottom SCC of $N_\alpha$, which is a semi-T-component of $N$ by Proposition~\ref{prop:botSemiT}.
\ENDPF

Now we recall the standard definition of T-components for general nets. They are special cases of semi-T-components, which leads us to introduce the notion of \emph{proper} semi-T-components. It will turn out that proper semi-T-components do not occur in well-formed free-choice nets; consequently, these nets are covered by T-components.

\DEF{def:Tcomp}{T-component, proper semi-T-component; of Type I and II}
    A~subnet $Y$ of a net $N$ is a~\emph{T-component of} $N$ 
    if $T_Y \neq \emptyset$ and the following conditions hold:
    \begin{enumerate}
        \item
        $Y$ is strongly connected;
        \item
        $|\post{s} \cap T_Y| = 1 = |\pre{s} \cap T_Y|$ for each $s \in S_Y$;
        \item
        $\post{T_Y} = S_Y = \pre{T_Y}$.
    \end{enumerate}

    A \emph{semi-T-component} $Y$ of $N$ is \emph{proper} if it is not a T-component. 
	By the definition of semi-T-components, such $Y$ must be of Type I, Type II, or both, where these types are defined as follows:
	\begin{itemize}
		\item \emph{Type I} (informally called ``excessive arc''): 
			$|\pre{s} \cap T_Y| \geq 2$ for some $s \in S_Y$ (that is, 
			$s$ has more than one input arc in $Y$);
		\item \emph{Type II} (``inbound arc''): $\pre{T_Y} \setminus
			\post{T_Y}\neq\emptyset$, that is, there is an arc $(s,t)\in F_N$
			such that $s\notin S_Y=\post{T_Y}$ and $t\in T_Y$.
	\end{itemize}
\ENDDEF

\begin{example}
	In Figure~\ref{fig:five_clusters_and_bott_scc}, the upper panel
	highlights a proper semi-T-component of
	Type II but not Type I.
	The lower panel shows  a~proper semi-T-component $Y$ that is of both Type I 
	(since $|\pre{s_{11}} \cap T_Y| \geq 2$)
	and Type II (since $s_{32}\in\pre{T_Y} \setminus \post{T_Y}$).
\end{example}

\PROP{prop:noProperSemiT}{Semi-T-components of well-formed free-choice net are T-components}
	If $N$ is a~well-formed free-choice net, then each semi-T-component of $N$ is a T-component.
\ENDPROP
\PF
	Let $N$ be a well-formed free-choice net and
	$Y$ a semi-T-component of $N$; we aim to show that $Y$ is in fact a
	T-component. 
	W.l.o.g.\ we assume that $N$ is
	strongly connected (recall Theorem~\ref{th:wfImpSc}).
	As $T_Y \neq \emptyset$, the net $N$ is not a single place; hence, the strong connectivity of $N$ ensures that every cluster in $N$ contains at least one transition.
	We choose a directed allocation $\alpha:\clusters_N\to T_N$ such that 
	$Y$ is the bottom SCC of $N_\alpha$, which is possible by
	Proposition~\ref{prop:botSemiT}.
		
	Let $M_0$ be a live and
	bounded marking of $N$ (which exists, since $N$ is well-formed). We
	note that $(N_\alpha,M_0)$ is deadlock-free. Indeed,
	if there were an execution $M_0\gt{\rho}M$ of $N_\alpha$ where $M$ is
	a~dead
	marking (meaning that each
	cluster $C$ contains a place $s$ with $M(s)=0$), then 
	$M_0\gt{\rho}M$, as an execution of $N$, would contradict the assumption  
	that $(N,M_0)$ is live.
	We can thus 
	\begin{center}
		fix an~arbitrary infinite execution $M_0 \gt{\sigma}$\ \ of $N_\alpha$.
	\end{center}
	Let $T^\sigma_\infty$ denote the nonempty set of transitions occurring
	infinitely often in $\sigma$.
	Due to the boundedness of $M_0$, if $t\in T^\sigma_\infty$ and
	$\post{t}\cap S_C\neq\emptyset$ for some cluster $C$, then the
	transition $\alpha(C)$ must also belong to $T^\sigma_\infty$ 
	(otherwise tokens would accumulate indefinitely in $S_C$ throughout the
	execution $M_0\gt{\sigma}$).
	Hence, if $t\in T^\sigma_\infty$ and there is a path
	from $t$ to $t'$ in $N_\alpha$, then $t'\in T^\sigma_\infty$.
	Our construction of $\alpha$ as an allocation directed to the bottom
	SCC $Y$ thus guarantees that $T_Y\subseteq
	T^\sigma_\infty$.

	Let $\sigma'$ arise from $\sigma$ by omitting all transitions from
	$T_Y$, that is, $\sigma'=\sigma\restr{T_N\setminus T_Y}$. Since
	$Y$ is a~bottom SCC of $N_\alpha$,  $M_0 \gt{\sigma'}$ is
	also an execution of $N_\alpha$ (by
	Proposition~\ref{prop:omitBotSCC}). 
	Furthermore,
	$\sigma'$ must be finite; otherwise, we would derive 
	$T_Y\subseteq T^{\sigma'}_\infty$ as before, which contradicts the
	definition of $\sigma'$.
	We thus obtain 
	\begin{center}
		$T^\sigma_\infty=T_Y$. 
	\end{center}
	This excludes the possibility that $Y$ is a proper semi-T-component of Type II
	(inbound arc). Indeed, if there were a place $s\in
	\pre{T_Y}\setminus \post{T_Y}$, then in the execution  $M_0
	\gt{\sigma}$,
	the marking of $s$
	would be infinitely often decreased (by transitions from $T_Y$) while only finitely many times
	increased (by transitions in $\sigma'$), which contradicts the
	boundedness of $(N,M_0)$.
	Hence, $\post{T_Y}=S_Y=\pre{T_Y}$.

	It remains to exclude that $Y$ is a proper semi-T-component of Type I.
	For the sake of contradiction, let 
	$s_0$ be a place in $S_Y$ and $t_1,t_2$ be
	two distinct transitions in $T_Y$ such that
	$\{t_1,t_2\}\subseteq\pre{s_0}$. In this case, 
	we consider a~simple cycle in the strongly connected subnet
	$Y$ that contains the arc $(t_1,s_0)$.
	Such a cycle exists because there is a 
	simple path from $s_0$ to $t_1$ in $Y$; moreover, the cycle cannot contain 
	the arc $(t_2, s_0)$. 
	Let $S'\subseteq S_Y$ and $T'\subseteq T_Y$ denote the sets of places and
	transitions in this cycle, respectively.

	For each $t\in T'$, we have $|\pre{t}\cap S'|=1\leq |\post{t}\cap
	S'|$.
	For each  $t\in T_Y\setminus T'$, we have $|\pre{t}\cap S'|=0\leq |\post{t}\cap S'|$
	(recall that each cluster contains at most one transition from $T_Y$).
	Crucially,  for the transition $t_2 \in T_Y$, we have a strict inequality:
	if $t_2 \in T'$, then $|\pre{t_2} \cap S'| = 1 < 2 \leq |\post{t_2} \cap S'|$ 
	(due to the arc $(t_2,s_0)$ and the arc $(t_2,s)$ belonging to the
	cycle, where $s\neq s_0$); 
	otherwise, $|\pre{t_2} \cap S'| = 0 < 1 \leq |\post{t_2} \cap S'|$.

	Since  $T_Y=T^\sigma_\infty$, the strict imbalance at $t_2$ ensures that the sum of tokens on the set $S'$ 
	increases above any bound in the execution $M_0 \xrightarrow{\sigma}$, 
	which contradicts the boundedness of $(N, M_0)$.
\ENDPF

The following lemma summarises the crucial points of the previous
propositions and proofs. (In this paper, we regard a lemma as a
crucial step towards the main theorem; it is typically built upon
several supporting propositions.)

\LEM{lem:Tcover}{T-coverability of well-formed free-choice nets}
	Every well-formed free-choice net $N$ is covered by
	T-components, that is,
	each $t\in T_N$ is in $T_Y$ for some~T-component $Y$ of $N$. 
	A set of such T-components covering $N$ can be constructed in polynomial
	time.
	Moreover, there is no proper semi-T-component in $N$.
\ENDLEM
\PF
	The first claim (the T-coverability theorem) and the last claim follow from the facts
	that every strongly connected free-choice net $N$ is covered by
	semi-T-components (Proposition~\ref{prop:semiTcover}), and that 
	each semi-T-component is a T-component if $N$ is, moreover, well-formed (Proposition~\ref{prop:noProperSemiT}).

	The construction of an allocation directed to $t_0\in T_N$, and of the
	bottom SCC containing $t_0$, in the
	proof of Proposition~\ref{prop:semiTcover}, is clearly polynomial in
	the size of the net $N$. This implies the second claim of the lemma.
\ENDPF

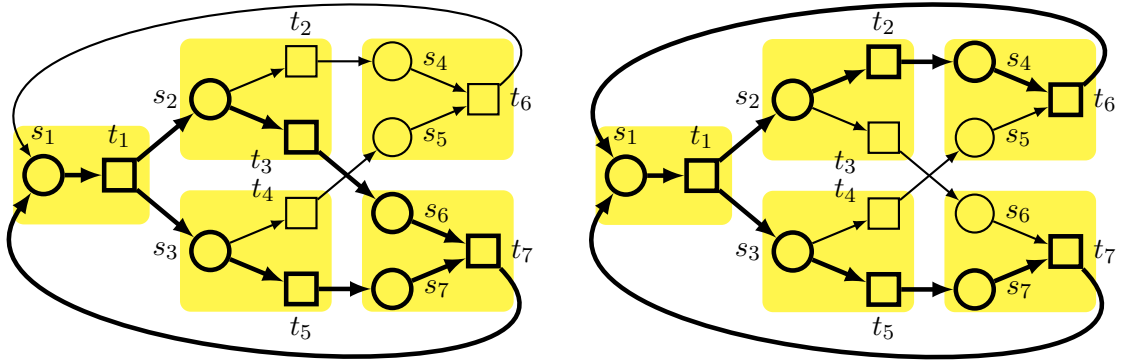
\begin{figure}[t]
    \centering
    \makebox[\linewidth][l]{
        \hspace*{-1cm}
        \begin{minipage}[t]{0.48\linewidth}
            \vspace{0pt}
            \centering
            \input{figures/t_covered.tex}
        \end{minipage}
		\hspace*{0cm}
        \begin{minipage}[t]{0.48\linewidth}
            \vspace{0pt}
            \centering
            \input{figures/proper_covered.tex}
        \end{minipage}
    }
	\vspace{-1cm}
    \caption{A non-well-formed net coverable by T-components (one of them is highlighted in bold on the left, the second one is symmetric) and
	a~proper semi-T-component (highlighted in bold on the right) of both Type I
	(excessive input arc at $s_1$) and Type II (inbound arcs $(s_5,t_6)$ and $(s_6,t_7)$).}
    \label{fig:not_wf_covered}
\end{figure}

\begin{example}
	Figure~\ref{fig:not_wf_covered}	shows a net $N$ that is covered by
	T-components (as well as by S-components defined below). The net has
	also a proper semi-T-component, which indicates that it is not
	well-formed.
\end{example}

The T-components have a natural counterpart: the S-components.
For defining S-components, as well as semi-S-components,
the concept of 
\emph{reverse-dual nets} is convenient; it consists in exchanging the roles of
places and transitions while 
reversing the arcs between them.

\DEF{def:rd}{Reverse-dual net $\rd(N)$}
	Let $N = (S,T,F)$ be a net. The \emph{reverse-dual net of} $N$ is
	the net $\rd(N)=(T,S,F^{-1})$.
\ENDDEF

The next proposition 
highlights some simple observations:

\PROP{prop:rdProp}{Properties of reverse-duality}
	Let $N$ be a net. Then:
	\begin{enumerate}[label=\arabic*)]
		\item $\rd(\rd(N))=N$.
		\item If $N$ is strongly connected, then so is $\rd(N)$.
		\item If $N$ is free-choice, then so is $\rd(N)$; moreover, if $C$ is a cluster of $N$, then $\rd(C)$ is a cluster of $\rd(N)$.
	\end{enumerate}
\ENDPROP

We could use Proposition~\ref{prop:ScorrespT} (below) as a definition of
``S-notions'', but for better transparency we provide an explicit
definition, by which    Proposition~\ref{prop:ScorrespT} becomes a
straightforward observation.

\DEF{def:Snotions}{Semi-S-component, S-component, proper
semi-S-component; of Type I and II}
	A~subnet $X$ of a net $N$ is a~\emph{semi-S-component of} $N$ 
	if $S_X \neq \emptyset$ and the following conditions hold:
	\begin{enumerate}
		\item
		$X$ is strongly connected (which implies 
		$T_X\subseteq \pre{S_X}\cap\post{S_X}$);
		\item
		$|\pre{t} \cap S_X|=1$ for each $t\in
		T_X$ (note that $|\post{t} \cap S_X|\geq 1$ follows from strong connectivity);
		\item
		$\pre{S_X}\subseteq T_X$ (hence  
		$\pre{S_X}=T_X\subseteq\post{S_X}$).
	\end{enumerate}
	If b) is strengthened to 
	``$|\pre{t} \cap S_X|=1=|\post{t} \cap S_X|$ for each $t\in T_X$'' and 
	c) is strengthened to ``$\pre{S_X}=T_X=\post{S_X}$'', then $X$ is an
	\emph{S-component}.

	A \emph{semi-S-component} $X$ \emph{of} $N$ is \emph{proper} if it is
	not an S-component. 
	Such $X$ must be of Type I, Type II, or both, where these types are defined as follows:
	\begin{itemize}
		\item \emph{Type I} (informally called ``excessive arc''): 
			$|\post{t} \cap S_X| \geq 2$ for some $t \in T_X$ (that is, 
	$t$ has more than one output arc in $X$);
		\item \emph{Type II} (``outbound arc''): $\post{S_X} \setminus
			\pre{S_X}\neq\emptyset$, that is, there is an arc $(s,t)\in F_N$
			such that $s\in S_X$ and $t\notin T_X=\pre{S_X}$.
	\end{itemize}
\ENDDEF

\PROP{prop:ScorrespT}{S-notions and T-notions correspond via reverse-dual nets}
	A subnet $X$ of a net $N$ is an \emph{S-component} (a
	\emph{semi-S-component}, a \emph{proper semi-S-component}, \emph{of
	Type I} and/or \emph{Type II}) of $N$ if and only if
	$\rd(N)[S_X\cup T_X]$ is a T-component (a semi-T-component, a proper
	semi-T-component, of Type I and/or Type II) of $\rd(N)$. 
\ENDPROP

Using Propositions~\ref{prop:rdProp} and~\ref{prop:ScorrespT}, 
we can readily derive Propositions~\ref{prop:topSemiS} and~\ref{prop:semiScover},
the analogues of 
Propositions~\ref{prop:botSemiT} and~\ref{prop:semiTcover},
after introducing the notion of place-allocations,
an analogue of (transition-)allocations.

 \DEF{def:placealloc}{Place-allocation $\beta$, induced subnet
    $N_\beta$, co-directed place-allocation}
Given a~free-choice net $N$ where each cluster contains at least one
place,
a \emph{place-allocation} is a~function
$\beta: \clusters_N \rightarrow S_N$ such that $\beta(C)\in S_C$ for
each cluster $C$.

Given a place-allocation $\beta$, we denote by $N_\beta$ the subnet
induced by the set $T_N$ of all transitions of $N$ and the set
$S_{N_\beta}=\{\beta(C)\mid C\in\clusters_N\}$ of all chosen
places.

A \emph{place-allocation} $\beta$ is \emph{co-directed from} a \emph{node} $u \in S_N
\cup T_N$ if for every node $v$ of $N_\beta$ there is a~path from $u$
to $v$ in
$N_\beta$ (which implies that $u\in S_{N_\beta}\cup T_{N}$). It is a
\emph{co-directed place-allocation} if it is co-directed from
some node (that is, if $N_\beta$ has exactly one top SCC).
\ENDDEF

\PROP{prop:topSemiS}{Top SCCs of $N_\beta$ coincide with semi-S-components of $N$}
Given a~free-choice net $N$ where each cluster contains at least one
place, a subnet $X$ is a semi-S-component of $N$ if and only if
it is a top SCC of $N_\beta$ for some place-allocation $\beta$.
If $N$ is, moreover, strongly connected, then any semi-S-component
is the top SCC of $N_\beta$ for some \emph{co-directed} place-allocation $\beta$.
\ENDPROP

\PROP{prop:semiScover}{Strongly connected free-choice net is covered
by semi-S-components}
       Every strongly connected free-choice net $N$ is covered by
       its semi-S-components; that is, each $s\in S_N$ belongs to
       $S_X$ for some
       semi-S-component $X$ of $N$. 
  \ENDPROP

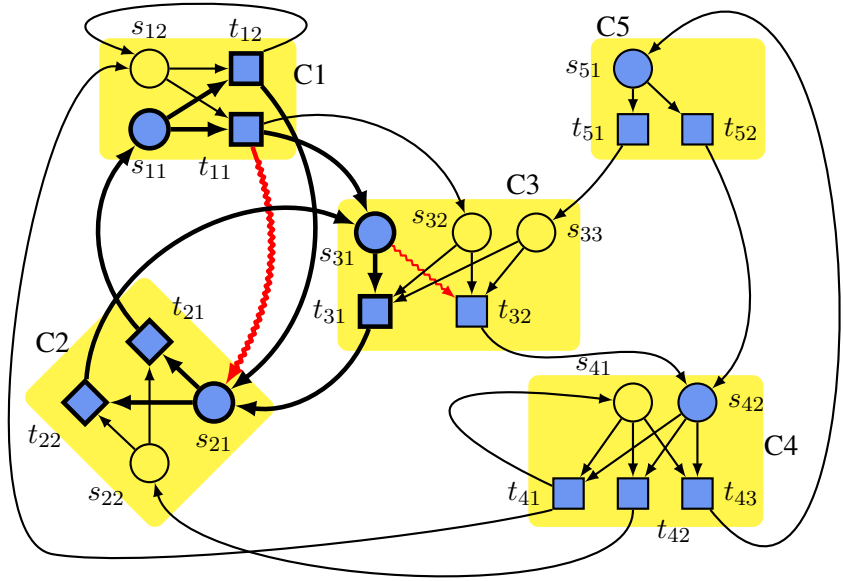
\begin{figure}[t]
    \centering
	\vspace{-1.6cm}
	\input{figures/five_clusters_dual.tex}
	\vspace{-1.2cm}
    \caption{A reverse-dual net $\rd(N)$ of the net $N$ from
	Figure~\ref{fig:five_clusters_and_bott_scc} (top panel).
	The blue subnet depicts the net $N_{\beta}$ where $\{s_{11}, s_{21}, s_{31}, s_{42}, s_{51}\}$
	is the set of $\beta$-allocated places.
	Its top SCC (depicted in bold) is a~semi-S-component of Type I (due to the ``excessive arc'' $(t_{11}, s_{21})$)
	and also of Type II (due to the ``outbound arc'' $(s_{31}, t_{32})$).}
    \label{fig:five_clusters_dual}
\end{figure}

The above discussed duality is illustrated in Figure~\ref{fig:five_clusters_dual}.

It is now tempting to formulate S-coverability of well-formed free-choice
nets as an analogue of T-coverability (Lemma~\ref{lem:Tcover}).
However, we cannot readily 
confirm that proper semi-S-components 
do not exist in well-formed free-choice nets; that is, an analogue of 
Proposition~\ref{prop:noProperSemiT} is not immediate. 
Nevertheless, 
the following proposition is immediate.

\PROP{prop:noProperSemiS}{Semi-S-components of $N$ are S-components
whenever $\rd(N)$ well-formed}
If $N$ is a free-choice net such that its reverse-dual net $\rd(N)$ is well-formed, then
each semi-S-component of $N$ is an S-component. This implies that $N$
is covered by S-components (that is, each $s\in S_N$ is in $S_X$ for
some S-component $X$ in $N$).
\ENDPROP

The issue of S-coverability will thus be settled by the following lemma.

\LEM{lem:feasiblewf}{Free-choice net with no proper semi-S, covered by
S-components, is well-formed} 
If a free-choice net $N$ is covered by S-components and there are no
proper semi-S-components in $N$, then $N$ is well-formed.
\ENDLEM

We postpone the proof of the lemma, first noting the main consequence
that implies the well-known \emph{coverability theorems} and \emph{duality 
theorem}~(cf.~\cite{de95}).

\THM{thm:wfChar}{Characterisation of well-formed free-choice nets by semi-components}

For a free-choice net $N$, the following conditions  are equivalent: 
        \begin{enumerate}
        \item
            $N$ is well-formed,
    \item
$N$ is covered by T-components and
	    there are no proper semi-T-components in $N$,
    \item
 $N$ is covered by S-components and
	    there are no proper semi-S-components in $N$,
 \item
	 the reverse-dual net $\rd(N)$ is well-formed.
	\end{enumerate}
\ENDTHM
\PF
Let $N$ be a free-choice net.
Recall that $\rd(N)$
is a free-choice net as well, by
Proposition~\ref{prop:rdProp}.

a)$\Rightarrow$b):
If a) holds, that is, $N$ is well-formed, then b) holds by
Lemma~\ref{lem:Tcover}.  

b)$\Rightarrow$d):
If b) holds, then $\rd(N)$ is covered by
S-components and there are no proper semi-S-components in $\rd(N)$, by
the assumption b) and
Proposition~\ref{prop:ScorrespT}. Hence,
$\rd(N)$ is well-formed by Lemma~\ref{lem:feasiblewf}; that is, d)
holds.

d)$\Rightarrow$c): If  $\rd(N)$ is well-formed, then $\rd(N)$ 
is covered by
T-components and there are no proper semi-T-components in $\rd(N)$, 
by Lemma~\ref{lem:Tcover}. Hence,  $\rd(\rd(N))$ is covered by
S-components and there are no proper semi-S-components in $\rd(\rd(N))$, by
Proposition~\ref{prop:ScorrespT}. 
Since  $\rd(\rd(N))=N$, c) is established.

c)$\Rightarrow$a): This follows by Lemma~\ref{lem:feasiblewf}.
\ENDPF

The proof of Theorem~\ref{thm:wfChar} will thus be finished by proving
Lemma~\ref{lem:feasiblewf}. 

\paragraph{Proof of Lemma~\ref{lem:feasiblewf}.}
We consider a free-choice net $N_0$ such that  $N_0$ is covered by S-components and there are no
proper semi-S-components in $N_0$.  We aim to demonstrate that $N_0$ is well-formed.

We start by  noting the \emph{token-conservation property of
S-components}, for any net $N$.

\PROP{prop:SConserv}{Token conservation in S-components, and structural boundedness}
If $X$ is an S-component of a net $N$, the sum of tokens in the places 
of $X$ remains constant during any execution of $N$.
\\
Consequently, if a net $N$ is covered by S-components, then $N$ is 
structurally bounded (that is, $(N, M_0)$ is bounded for every initial 
marking $M_0$).
\ENDPROP
\PF
	The first part is clear by recalling that for any S-component $X$ of
	$N$ we have:  
	$|\pre{t}\cap S_X|=|\post{t}\cap S_X|=1$ for each $t\in T_X$,
	and  $|\pre{t}\cap S_X|=|\post{t}\cap S_X|=0$
	for each $t\in T_N\setminus
	T_X$ (since $\pre{S_X}=T_X=\post{S_X}$).

	The consequence follows by observing that for 
	a fixed set $\text{SC}$ of S-components that cover the net $N$, and 
	for all
	$M\in\rset{M_0}$ we have
	\begin{equation*}
	\sum_{s\in S_N} M(s)\leq
	\sum_{X\in\text{SC}}\sum_{s\in
	S_X}M(s)=\sum_{X\in\text{SC}}\sum_{s\in S_X}M_0(s).
	\end{equation*}
\ENDPF

Hence, the considered free-choice net $N_0$ is \emph{structurally
bounded}. 
The following proposition is a~crucial step for 
establishing \emph{structural liveness} (the existence of at least one live
marking) of $N_0$. Recall that a marking $M$ of a net $N$ is a DL-marking if
each transition of $N$ is either dead or live at $M$ and at least one
transition is dead.

\PROP{prop:MDLtopSCC}{DL-markings, and semi-S-components with no tokens}
Let $M$ be a DL-marking of a free-choice net $N$.
Then there exists a semi-S-component $X$ of $N$,
such that $M\restr{S_X}=\mathbf{0}$.
\ENDPROP
\PF
Given a DL-marking $M$ of a free-choice net $N$, 
we present $T_N$ as the disjoint union $T\dead \sqcup T\live$ 
of the sets of transitions that are dead and live at $M$, 
respectively, where $T\dead \neq \emptyset$.
For each cluster $C\in\clusters_N$ with $T_C\neq\emptyset$, we thus have
either $T_C\subseteq T\dead$ or $T_C\subseteq T\live$. 
Let
\begin{equation*}
\clusters\dead = \{C \in \clusters_N \mid T_C\subseteq
	T\dead \text{ and } T_C\neq\emptyset\}
\end{equation*}
be the set of \emph{dead clusters} at $M$.
(We ignore the possible cluster collecting all places $s\in S_N$ with
$\post{s}=\emptyset$.)

We define a \emph{partial place-allocation}
$\beta\colon\clusters\dead\to S_N$ such that, for
each $C\in\clusters\dead$, 
\begin{center}
the place $s = \beta(C) \in S_C$ 
satisfies $M(s)=0$ and $\pre{s}\subseteq T\dead$.
\end{center}
	This choice is indeed possible:
if for each $s\in S_C$ we had either $M(s)\geq 1$ or $\pre{s}\cap
T\live\neq\emptyset$, then $C$ would not be dead at $M$, since it
could eventually become enabled by firing transitions from $T\live$.

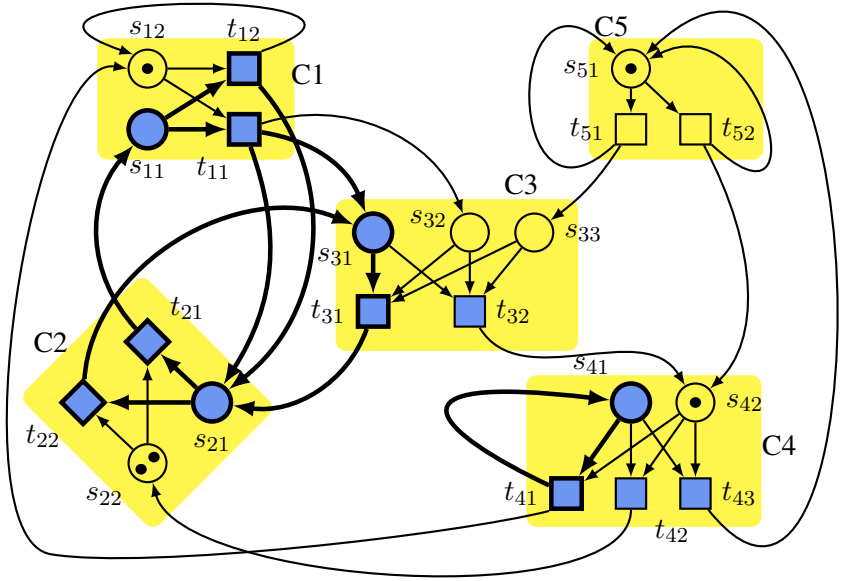
\begin{figure}[t]
    \centering
	\vspace{-1.6cm}
    \input{figures/dl_marking_dual.tex}
	\vspace{-1.2cm}
    \caption{A free-choice net with a DL-marking $M$.
	Cluster C5 is live, the other clusters are dead.
	The blue subnet $D$ corresponds to
	a partial place-allocation selecting places
	in dead clusters that are unmarked and
	have no live input transitions.
	The top two SCCs of $D$ (in bold) are unmarked semi-S-components
	(that are in this case proper; one of Type I \& II, the other of Type II only).}
	\label{fig:dl_marking_dual}
\end{figure}

Let $D$ be the subnet of $N$ induced by the set of places
$S_D=\{s\mid s=\beta(C)\text{ for some } C\in\clusters\dead\}$ 
and the set of transitions $T_D=T\dead$. 
(See the illustration in Figure~\ref{fig:dl_marking_dual}.)
Let $X$ be a top SCC 
of the subnet $D$ (which might be a single place, but not a 
single transition).

Hence $X$ is
strongly connected, $S_X\neq\emptyset$,
and $|\pre{t}\cap S_X|=1$ for each $t\in T_X$ (since each cluster of
$D$, and thus also of its strongly connected subnet $X$, contains precisely
one place).

Moreover,
$\pre{S_X}\subseteq T_X$ (where the notation ``$\pre{}$'' refers to
the net $N$),
since $\pre{S_X}\subseteq T_D=T\dead$ and 
$X$ is a \emph{top} SCC of $D$ (thus having no incoming arcs in $D$).
The subnet $X$ is thus a semi-S-component of $N$ with $M(s)=0$ for all $s\in
S_X$, the existence of which we aimed to establish.
\ENDPF

\begin{remark}
Proposition~\ref{prop:MDLtopSCC} thus confirms the motivating
observation made in
	Remark~\ref{rem:bottomYtopX}, since the semi-S-component $X$
	in the proof is a top SCC of $N_{\beta}$ for some
	place-allocation $\beta$. Note that 
the partial place-allocation 
	$\beta$ constructed in the proof can be extended to a (total)
	place-allocation if every cluster contains at least one place.
\end{remark}

We continue the plan to show that the net $N_0$ under consideration is well-formed; its structural boundedness has already been established. The following proposition establishes the structural liveness of $N_0$, relying on the fact that $M_0$ is live if and only if no DL-marking is reachable from $M_0$ (Proposition~\ref{prop:dlmarking}). This proposition thus completes the proof of Lemma~\ref{lem:feasiblewf}.

Recall that
a proper semi-S-component $X$ of Type II has at least one outbound arc
$(s,t)$ with $s\in S_X$, $t\notin T_X$, and $\post{t}\cap S_X=\emptyset$.

\PROP{prop:noProperSwf}{Free-choice net with no proper
semi-S of Type II 
is structurally live}
If a free-choice net $N$ has no proper semi-S-component of Type II
(that is, one with an outbound arc), then any
marking $M_0$ of $N$ that places at least one token in each semi-S-component 
(for instance, the marking $M_0$ where $M_0(s)=1$ for all $s \in S_N$) is live.
\ENDPROP
\PF
For any marking $M_0$ satisfying the assumption, no DL-marking $M$ is
reachable. Indeed, the semi-S-component $X$ guaranteed for such an $M$ by
Proposition~\ref{prop:MDLtopSCC} would necessarily be either an S-component
or a proper semi-S-component of Type I, but not of Type II; for such
$X$,
each transition $t\in \post{S_X}$ satisfies $|\pre{t}\cap
S_X|=1\leq|\post{t}\cap S_X |$. 
However, this semi-S-component $X$ would have lost all its tokens during the
hypothetical execution from $M_0$ to $M$, which is impossible.
It follows that $M_0$ is live, by Proposition~\ref{prop:dlmarking}.
\ENDPF

\begin{remark}
	Proposition~\ref{prop:noProperSwf} is more general than what is strictly required to conclude the proof of Lemma~\ref{lem:feasiblewf}.
In particular, it highlights the significance of
proper semi-S-components of Type II and will also be utilised in
Appendix~\ref{sec:alg}.
\end{remark}

\section{Concluding remarks}
\label{sec:remarks}

Semi-T-components and semi-S-components are the two central concepts supporting the theory developed in Section~\ref{sec:main}. Given a free-choice net $N$, the sets $S_X$ of semi-S-components $X$ (satisfying $\pre{S_X} \subseteq \post{S_X}$) correspond to minimal siphons. These have traditionally played a crucial role in classical proofs, such as the proof of Commoner's Theorem (see Appendix~\ref{commoner.app}) and the proof of S-coverability for well-formed free-choice nets (see, e.g., Chapters 4 and 5 in~\cite{de95}, where minimal siphons are essential, or Theorem 5.34 in~\cite{DBLP:books/sp/BestD24}, where Commoner's Theorem is employed extensively).

Dually, semi-T-components $Y$ correspond to what might be termed minimal T-traps, that is, transition subsets $T_Y \subseteq T$ satisfying $\post{T_Y} \subseteq \pre{T_Y}$. While we utilize these structures to derive the T-coverability of well-formed free-choice nets, neither T-traps nor semi-T-components appear to have played such a prominent role in the history of T-coverability results as they do in the present paper. The archetypal proof in~\cite{de95} employs T-invariants and an exchange lemma, whereas the original proof by Hack~\cite{hack72} resorts to a reduction technique that is semi-formally justified.

A polynomial-time algorithm for deciding the well-formedness of strongly connected free-choice nets,
which is quite different from Algorithm~\ref{alg:wf},
was proposed in Chapter~6 of \cite{de95}.
This algorithm is based on the Rank Theorem for free-choice nets,
which can be reduced to solving systems of linear equations.

Another algorithm for deciding the well-formedness of free-choice nets, relying on a modified version
of the Rank Theorem, was described by Kemper and Bause in~\cite{DBLP:conf/apn/KemperB92} and later improved
in~\cite{DBLP:conf/apn/Kemper93} and~\cite{kemper_cover}.
This algorithm achieves an overall time complexity of \(O(|S||T|^2)\)
for deciding the well-formedness of (strongly connected) free-choice nets.
To the best of our knowledge, this is currently the most efficient algorithm known.

Barkaoui and Minoux proposed a polynomial-time algorithm for deciding the liveness of bounded free-choice nets
\cite{DBLP:conf/apn/BarkaouiM92}.
This was later improved and extended by Barkaoui, Couvreur and
Dutheillet
in \cite{DBLP:conf/apn/BarkaouiCD95}; with respective complexity
\(O(|S||T||F|)\).
Our Algorithm~\ref{alg:wf} in Appendix~\ref{sec:alg} adopts a dual approach to theirs,
being based on semi-T-components. We also obtain the bound
$O(|S||T||F|)$ (which is in $O(|S|^2|T|^2)$, since $|F|\leq 2|S||T|$).

As part of follow-up research, we plan to explore whether this approach can
be optimised to match the bounds achieved by algorithms based on the Rank
Theorem. It is worth noting, however, that the quality of such an algorithm
does not lie in its time complexity alone, but also in the usefulness of its
output. Our algorithm returns a cover by T-components in the positive case
and, in the negative case, a proper semi-T-component witnessing the failure;
algorithms that decide well-formedness by rank computations do not naturally
yield such a witness.

\section*{Acknowledgement} 

The authors are grateful to the reviewers for their insightful comments.

\newpage
\appendix

\section{A polynomial algorithm deciding well-formedness of free-choice nets}
\label{sec:alg}

We propose 
Algorithm~\ref{alg:wf}, with
Algorithm~\ref{alg:semiT} as its subprocedure. The restriction to
strongly connected nets in the input of
Algorithm~\ref{alg:wf} is harmless, due to Theorem~\ref{th:wfImpSc}
and the availability of linear-time algorithms for computing the SCCs of a graph
(in time $O(|V|+|E|)$ for a graph $G=(V,E)$, hence in time
$O(|S|+|T|+|F|)$ for a net $N=(S,T,F)$).
Correctness relies primarily on Theorem~\ref{thm:wfChar}. 
Specific arguments are presented via inline comments in the algorithms, referring to the points summarised in the following proposition.

\PROP{prop:algcorrect}{Arguments for correctness of
Algorithms~\ref{alg:wf} and \ref{alg:semiT}}
\begin{enumerate}[label=\arabic*)]
	\item		
		If a free-choice net $N$ is covered by T-components, then $N$ is
		well-formed if and only if there exists no proper semi-T-component of
		Type II in $N$. 
	\item
If $Y$ is a subnet of a net $N$ and $t \in T_N \setminus T_Y$, then
		$Y$ is a semi-T-component in $N$ if and only if it is
		a semi-T-component in the net $N_t$ obtained from $N$
		by removing the transition $t$.
	\item
		If $Y$ is a subnet of a net $N$ and 
		$(\{s\}\cup\pre{s})\cap (S_Y\cup T_Y)=\emptyset$, then 
		$Y$ is a semi-T-component in $N$ if and only if it is
		a semi-T-component in the net $N_s$ obtained from $N$
		by removing the nodes in $\{s\} \cup \pre{s}$.
		
	\item
		In a net $N$, there exists a proper semi-T-component of Type~II if and only if there is some $s \in \pre{T_Y} \setminus \post{T_Y}$, where $Y$ is a semi-T-component in the net $N_s$ obtained from $N$ by removing the nodes in $\{s\} \cup \pre{s}$.

\end{enumerate}
\ENDPROP

\PF
1. Let $N$ be a free-choice net covered by T-components. Then the
free-choice net
$\rd(N)$ is covered by S-components; thus, $\rd(N)$ is structurally
bounded (by Proposition~\ref{prop:SConserv}).
Hence, $\rd(N)$ is well-formed if and only if there is no proper semi-S-component of Type~II in~$\rd(N)$; specifically, the $\Leftarrow$ direction follows from Proposition~\ref{prop:noProperSwf}, while the $\Rightarrow$ direction is due to Theorem~\ref{thm:wfChar}(a)$\Rightarrow$(c).
Consequently, $N$ is well-formed if and only if 
there is no proper semi-T-component of Type II in $N$ 
(recall that $N$ is well-formed if and only if $\rd(N)$ is well-formed, by Theorem~\ref{thm:wfChar}(a)$\Leftrightarrow$(d)).

2.\ and 3.
The claims follow readily from an inspection of the conditions imposed on semi-T-components in Definition~\ref{def:semiTcomp}.
(The subnet $Y$ is strongly connected, $T_Y\neq\emptyset$, 
$|\post{s} \cap T_Y|=1$ for each $s\in S_Y$, $\post{T_Y}=S_Y$.) 

4.
Recall that a semi-T-component $Y$ in a net $N$ is a proper
semi-T-component of Type II if and
only if there exists $s \in \pre{T_Y} \setminus \post{T_Y}$; by
definition of semi-T-components (implying $\post{T_Y}=S_Y$), we have 
$(\{s\}\cup\pre{s})\cap (S_Y\cup T_Y)=\emptyset$.
The claim thus follows from item 3.
\ENDPF

\begin{algorithm}[H]
	\caption{Deciding if a free-choice net is well-formed}\label{alg:wf}
    
    \SetKwInOut{Input}{Input}
    \SetKwInOut{Output}{Output}

	\Input{A strongly connected free-choice net $N$.}
	\Output{\emph{Yes}, with a set $\mathit{TCover}$ of T-components that
	cover $N$, if $N$ is well-formed (wf); \\
	    \emph{No}, with a proper semi-T-component (psTc) $Y$ of $N$, if $N$ is not well-formed.}
    
    \BlankLine

$\mathit{TCover} \leftarrow \emptyset$ \;
$\mathit{covered} \leftarrow \emptyset$ \tcp*[r]{Set of transitions covered by $\mathit{TCover}$}
\BlankLine

\While{$\mathit{covered} \neq T_N$}{
    Select $t \in T_N \setminus \mathit{covered}$ \;
	Create an allocation $\alpha$ directed to $t$ \tcp*[r]{exists since $N$ is strongly connected} 
	Let $Y$ be the bottom SCC of $N_\alpha$\; 
	\tcp{Hence $t$
	belongs to $T_Y$, and $Y$ is an sTc in $N$
	(cf.~Prop.~\ref{prop:botSemiT}).}
		\eIf{$Y$ is a proper semi-T-component (a psTc)}{
	    \Return{{No}, $Y$}\label{line:FirstProperSemiT}\;
    }{
        $\mathit{TCover} \leftarrow \mathit{TCover} \cup \{Y\}$ \;
        $\mathit{covered} \leftarrow \mathit{covered} \cup T_Y$ \;
    }
}
\BlankLine
		\tcp{Construction of $\mathit{TCover}$ is complete,
		$N$ is covered by T-components.}
	\tcp{Hence $N$ is wf iff there is no psTc of Type II in
	$N$ (Prop.~\ref{prop:algcorrect}(1)).}
		\tcp{Search for  
		a psTc of Type II in $N$ follows.}

		\tcp{For each $s\in S_N$ check if
		$s\in
	\pre{T_Y}\setminus\post{T_Y}$ for some sTc $Y$, that is,}
	\tcp{if some $t\in T_{C(s)}$ belongs to some sTc $Y$ in the net
	$N_s$ of Prop.~\ref{prop:algcorrect}(4).}
	\tcp{This is impossible if
$s$ is a unique place in its cluster $C(s)$,}
		\tcp{since $N$ is strongly connected.}

		\BlankLine
	\ForEach{place $s \in S_{N}$ such that $|S_{C(s)}| > 1$}{		
	$T'\leftarrow T_{C(s)}\setminus\pre{s}$\;

	\If{$T'\neq\emptyset$}{
	Construct the subnet $N_s$ resulting from $N$ by removing the set of nodes $\{s\}
	\cup \pre{s}$\label{line:netNs} \;		
		\tcp{$N_s$ (is a free-choice net that) might not be strongly connected.}
	Call Algorithm~\ref{alg:semiT} for $N_s$ and $T'$\;
	\tcp{It returns an sTc $Y$ in $N_s$ intersecting
	$T'$, if any exists.}
	\If{\textnormal{Algorithm~\ref{alg:semiT}} returns
	$Y$ (hence $T_Y\cap T'\neq\emptyset$)\label{line:checksemi}}{
			\Return{{No},
			$Y$\label{line:SecondProperSemiT}}
\tcp*[r]{
			$s\in \pre{T_Y}\setminus\post{T_Y}$
			in $N$ (since $T'\subseteq T_{C(s)}=\post{s}$}
		\tcp{and $\pre{s}\cap T_Y=\emptyset$); 
			hence $Y$ is a psTc of Type II in $N$.}
    }
}
}
\BlankLine
		\tcp{No psTc of Type II exists in $N$; $N$ is well-formed.}		
\BlankLine
		\Return{{Yes}, $\mathit{TCover}$}\;
\end{algorithm}

\begin{algorithm}[!ht]
	\caption{Deciding if a transition set in an fc net intersects 
		some semi-T-component}\label{alg:semiT}
    
    \SetKwInOut{Input}{Input}
    \SetKwInOut{Output}{Output}
    \SetKwBlock{Repeat}{repeat}{}

		\Input{A (general) free-choice net $N$ and a set
		$T_0\subseteq T_N$.}
	\Output{A semi-T-component $Y$ such that
		$T_Y\cap T_0\neq\emptyset$, if such $Y$ exists; \\
            \emph{No} otherwise.}
		\tcp{A semi-T-component (sTc) $Y$ is ``admissible'' if $T_Y\cap
		T_0\neq\emptyset$.}
    \BlankLine

$\mathit{Net} \leftarrow N$ \tcp*[r]{$\mathit{Net}$ is a variable for successively reduced subnets of $N$}

\Repeat{
	Mark each $t\in T_{\mathit{Net}}$ as \emph{good}
	iff for each node $u\in\{t\}\cup\post{t}$ there is a path from $u$ to $T_0$\;
	\tcp{By a path from $u$ to $T_0$ we mean a path from $u$ to some $t \in T_0$.}
	\tcp{Note that for every admissible $Y$, every $t\in T_Y$ is
	good;}
	\tcp{indeed: any sTc $Y$ is strongly
	connected, and satisfies $\post{T_Y}=S_Y$.}

	\BlankLine
	\If{$T_0$ does not contain any good transition}{
		\Return{No}\label{line:t0nongood}\tcp*[r]{Clearly,
		no
		admissible sTc exists.}
	}
	
	\If{all $t\in T_{\mathit{Net}}$ are good}{
		\tcp{Note that each place $s\in S_{\mathit{Net}}$ with
		$\post{s}=\emptyset$ also satisfies
		$\pre{s}=\emptyset$.}

		$\mathit{Net} \leftarrow$ the subnet of $\mathit{Net}$ obtained by
		removing all isolated places, if there are any\;

		\tcp{Now each cluster of $\mathit{Net}$ has at least one transition.}
 
		Create an allocation
		$\alpha$ for $\mathit{Net}$ directed to $T_0$\;

		\tcp{That is, for every node $u$ of $\mathit{Net}_\alpha$ there is a path from $u$ to $T_0$.}

		\Return{a bottom SCC $Y$ of
		$\mathit{Net}_\alpha$}\;
	\tcp{$Y$ is an sTc by Prop.~\ref{prop:botSemiT}, and $T_Y\cap
	T_0\neq\emptyset$ due to the choice of $\alpha$}
}
		\BlankLine

		\tcp{Some $t\in T_0$ is good and there exist non-good
		transitions in $\mathit{Net}$:}
                
		\BlankLine

		$\mathit{Net} \leftarrow$ the subnet of $\mathit{Net}$ obtained by
		removing all non-good
		transitions\label{line:reducednet}\;
$T_0 \leftarrow$ the subset of $T_0$ obtained by
		removing all
		non-good transitions\label{line:reducedT0}\;
\BlankLine
\tcp{$\mathit{Net}$ remains a free-choice net,
cf.~Prop.~\ref{prop:subFree}(2).}
\tcp{The set of admissible sTcs remains unchanged, cf.~Prop.~\ref{prop:algcorrect}(2).}
}
\end{algorithm}

\bigskip
\THM{th:algo}{Correctness and Complexity of Algorithm~\ref{alg:wf}}
	\begin{enumerate}
		\item
			Algorithm~\ref{alg:wf} works correctly: it
			returns \emph{No}, with a proper
			semi-T-component $Y$ of the given strongly
			connected free-choice net $N=(S,T,F)$, if $N$ is not well-formed;
			it returns \emph{Yes}, with a T-component cover $\mathit{TCover}$, if $N$ is well-formed.
		\item
The time complexity of Algorithm~\ref{alg:wf}, 
assuming $N$ has more than one node,
			is in $O(|S||T||F|)$ (hence also in 
			$O(|S|^2 |T|^2)$ since $|F|\leq 2|S||T|$). 
	\end{enumerate}
\ENDTHM
\PF
a) The correctness of Algorithm~\ref{alg:wf} and its subprocedure 
			Algorithm~\ref{alg:semiT}  follows from the
			arguments presented in
			Proposition~\ref{prop:algcorrect} and the
			inline comments in the pseudocodes of the algorithms.

b)			
We first analyse the time complexity of Algorithm~\ref{alg:semiT}, and
then use it in analysis of  Algorithm~\ref{alg:wf}.
We utilise the fact that the time complexity of 
			\emph{breadth-first search} (BFS) is
			$O(|V| + |E|)$ on graph $G=(V,E)$, hence 
			 $O(|S| + |T| + |F|)$ in our case.

\emph{Complexity of Algorithm~\ref{alg:semiT}, given 
$N=(S,T,F)$ and $\emptyset\neq T_0\subseteq T$, is in $O(|T|\cdot(|S|
+ |T| + |F|))$.}
				\begin{itemize}
					\item \emph{Loop
						Iterations:} In each
						non-final
						iteration of the
						\texttt{repeat} loop
						(Lines~2--11),
at least one transition is identified as non-good and removed from the
						net (Lines~10--11),
						while some (good)
						transitions remain.
					Consequently, the \texttt{repeat} loop executes at most $|T|$ times.
					\item \emph{Cost per Iteration:} 
					\begin{itemize}
						\item
							Identifying good transitions (Line~3) requires checking reachability to $T_0$,
							which can be done in time $O(|S| + |T| + |F|)$ using a single BFS on the reversed flow relation $F^{-1}$.
\item
							If all
							remaining
							transitions
							are good
							(Line~6), the
							result of the
							aforementioned
							BFS allows an
							immediate
							construction
							of an
							allocation
							$\alpha$
							directed to
							$T_0$ (Line~8)
							by selecting
							the
							minimal-distance
							transition (to
							$T_0$) in
							each cluster.
						\item
							Extracting a bottom SCC of $\mathit{Net}_\alpha$ (Line~9)
							via standard linear-time SCC algorithms (such as Tarjan's algorithm)
							takes $O(|S| + |T| + |F|)$ time.
					\end{itemize}
				\end{itemize}

\medskip

\emph{Complexity of Algorithm~\ref{alg:wf}.} 
We consider a given strongly connected free-choice net $N=(S,T,F)$
with more than one node; hence $|S|+|T|\leq |F|$.
We analyse two sequential phases of the algorithm separately:

\emph{Phase 1: T-Component Coverability (Lines~1--11) in $O(|T||F|)$.}

			\begin{itemize}
					\item \emph{Loop Iterations:} 
In each iteration of the \texttt{while}
						loop (Line~3),
a semi-T-component $Y$ covering a so far uncovered transition is
					constructed (Line~6). 
The loop thus runs at most $|T|$ times.
					\item \emph{Cost per Iteration:} 
					\begin{itemize}
						\item
							Within the loop, constructing a directed allocation toward $t$ (Line~5)
							and finding
							the bottom SCC
							$Y$ of
							$N_\alpha$
							(Line~6) take
							$O(|S|+|T|+|F|)$ time
							(as described
							for Algorithm~\ref{alg:semiT}).
						\item
							Verifying whether $Y$ is a proper semi-T-component (Line~7)
							can be done in
							$O(|S|+|F|)$
							time via a
							pass over 
							the places,
							checking if
							$s\in S_Y$ has
							two
							ingoing arcs
							from $T_Y$
							(Type~I) or 
							$s\not\in S_Y$
							has an
							outgoing arc
							to $T_Y$
							(Type~II).
						\item
							Since 
							$|S|+|T|\leq
							|F|$, each
							iteration runs
							in $O(|F|)$
							time.
					\end{itemize}
			\end{itemize}
			
\emph{Phase 2: Type~II Proper Semi-T-Component Search (Lines~12--18)
in
$O(|S| |T| |F|)$}.
			\begin{itemize}
				\item \emph{Loop Iterations:} 
The \texttt{foreach} loop (Line~12) clearly executes at most $|S|$ times.
				\item \emph{Cost per Iteration:} 
				\begin{itemize}
					\item
						For each candidate place $s$, generating the reduced subnet
						$N_s$ (Line~15) takes $O(|F|)$ time, as it involves removing
						the nodes in $\{s\} \cup \pre{s}$ and their incident arcs.
					\item
						Invoking Algorithm~\ref{alg:semiT} on
						$N_s$ and $T'$ (Line~16)
						takes
						$O(|T|\cdot(|S|+|T|+|F|))$
						time, hence
						$O(|T||F|)$ (since
						$|S|+|T|\leq |F|$).
				\end{itemize}
			\end{itemize}
The complexity of Algorithm~\ref{alg:wf} thus is in $O(|S||T||F|)$
(hence in $O(|S|^2 |T|^2)$, since $|F| \leq 2 |S| |T|$).
\ENDPF

\begin{remark}
Due to duality, the algorithms can be adapted to use
semi-S-components instead of semi-T-components; this corresponds to 
operating on the reverse-dual net $\rd(N)$ instead of $N$.
\end{remark}

\section{Commoner's Theorem}
\label{commoner.app}

Given a net $N$, a set $Q\subseteq S_N$ is a \emph{trap} if $\post{Q}\subseteq\pre{Q}$,
and it is a \emph{siphon} if $\pre{Q}\subseteq\post{Q}$.
We will use the fact that the union of traps is itself a trap; thus,
every set 
$R\subseteq S_N$ contains a \emph{unique maximal trap} $Q\subseteq R$, namely the
union of all traps contained in $R$.
Additionally, we recall that if $Q$ is a trap such that $M_0\restr{Q} \neq \mathbf{0}$, then $M\restr{Q} \neq \mathbf{0}$ for all markings $M \in \rset{M_0}$.
On the other hand, if $R$ is a siphon and $M\restr{R}=\mathbf{0}$,
then all transitions from $\post{R}$ are dead at $M$ (since
$\pre{R}\subseteq\post{R}$).
 
We present Algorithm~\ref{alg:maxTrap}, a standard procedure for computing the maximal trap $Q$ inside a
set of places $R$ (which may, in particular, be a siphon). 
The algorithm also provides layers of ``leaking transitions'' for $R$.
We observe some
simple facts regarding these transitions
(Proposition~\ref{prop:maxTrapFacts}), which allow us to provide a
smooth proof of Commoner's Theorem.

\begin{algorithm}[H]
\caption{Construction of the maximal trap $Q$ inside
a set of places	$R$}\label{alg:maxTrap}

\SetKwInOut{Input}{Input}
\SetKwInOut{Output}{Output}
\SetKwBlock{Repeat}{repeat}{}

\Input{A net $N$ and a set of places $R \subseteq S_N$.}
\Output{The maximal trap $Q\subseteq R$.}

\BlankLine

$Q \leftarrow R$\tcp*[r]{$Q$ is a variable whose value will be returned at the end}
\Repeat{ 
	$T\exit \leftarrow \post{Q}\setminus \pre{Q}$\tcp*[r]{each $t\in T\exit$ causes that $Q$ is not
	a trap}
    \If{$T\exit = \emptyset$}{
	    \Return{$Q$}
	    \tcp*[r]{the returned $Q$ is a trap, since
	    $\post{Q}\subseteq \pre{Q}$} }
\tcp{no place $s\in\pre{T\exit}$ can be in the maximal trap inside $Q$}    
$Q \leftarrow Q \setminus \pre{T\exit}$\;
}
\end{algorithm}

Inspecting Algorithm~\ref{alg:maxTrap}, we can see the following.
Given a net $N$ and a set $R\subseteq S_N$, the algorithm creates a sequence
\[
   Q_0,\; T_1,\; Q_1,\; T_2,\; Q_2,\;\dots,\; T_m,\; Q_m
   \qquad(m\geq 0)
\]
of values of the variables $Q$ and $T\exit$, where $Q_0=R$ and, for each
$i\in[1,m]$, $T_i=\post{Q_{i-1}}\setminus\pre{Q_{i-1}}$ is nonempty and
$Q_i=Q_{i-1}\setminus\pre{T_i}$; the algorithm finishes by computing $T\exit=\emptyset$ and
returning $Q_m$, which is a trap since $\post{Q_m}\subseteq\pre{Q_m}$, and which
is in fact the maximal trap $Q\subseteq R$.

Each $T_i$ is a nonempty subset of $\post{R}$, and
$Q_i=R\setminus\bigcup_{j\in[1,i]}\pre{T_j}$ for $i\in[1,m]$.
Since $T_i\subseteq\post{Q_{i-1}}$, some place of $Q_{i-1}$ belongs to
$\pre{T_i}$; hence $R=Q_0\supsetneq Q_1\supsetneq\cdots\supsetneq
Q_m=Q$, which implies
that $m\leq|R\setminus Q|$.
The sets $T_1,T_2,\dots,T_m$ are pairwise disjoint: for $i<j$, every $t\in T_j$
has a place of $Q_{j-1}$ in its preset, whereas $Q_{j-1}\subseteq Q_i$ and
$\pre{T_i}\cap Q_i=\emptyset$.

\begin{figure}[h!]
    \centering
\input{figures/commoner.tex}
	\caption{Layers of leaking transitions
	$T_1=\{t_{11},t_{12}\}$, $T_2$, $T_3$ for a
	free-choice net; $t_0\in\post{R}\setminus T\leak$.}
	\label{fig:commoner}
\end{figure}
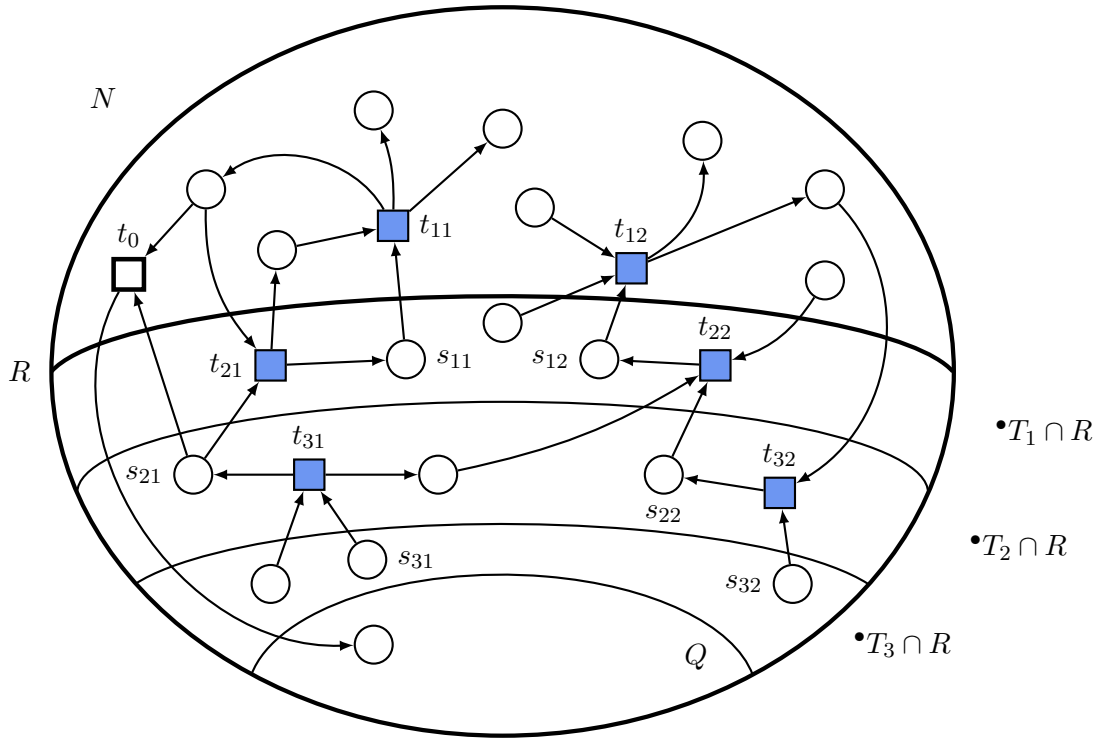

We view $T\leak=\bigcup_{i\in[1,m]} T_i$ as the set of \emph{leaking
transitions}. 
Each
leaking transition $t \in T\leak$ has a well-defined \emph{exit index}
$\mathit{ei}(t) \in [1, m]$, defined as $\mathit{ei}(t) = i$ if $t \in T_i$.
(See Figure~\ref{fig:commoner} for an illustration with $m=3$. For a
free-choice net, the condition $\pre{T_i} \cap \pre{T_j} = \emptyset$
holds for $i \neq j$, but this is not true in general.)

\PROP{prop:maxTrapFacts}{Leaking transitions as classified by
Algorithm~\ref{alg:maxTrap} for a set of places $R$}
Given a net $N$ and a set $R\subseteq S_N$, where $Q$ is the maximal trap
inside $R$,
the transitions from the
set
$T\leak=\bigcup_{i\in[1,m]} T_i$ satisfy:
\begin{enumerate}[label=\arabic*), ref=\arabic*]
\item\label{item:disjoint} 
	If $\mathit{ei}(t)=i{+}1$ (that is, $t\in T_{i+1}$,
		$i\in[0,m{-}1]$), then there exists some
	place $s_t$ satisfying $s_t\in \pre{t}\cap Q_i$
		(where $Q_i=R\setminus\bigcup_{j\in[1,i]}\pre{T_j}$);
		moreover $s_t\notin\bigcup_{j\in[1,i+1]}\post{T_j}$.
	\item
		$R\cap \pre{T\leak}=R\setminus Q$ (that is,
		$R\setminus \pre{T\leak}=Q$).
	\item 
		For each $t\in T\leak$ we have
		$\pre{t}\cap Q=\post{t}\cap Q=\emptyset$.
\end{enumerate}
		\ENDPROP

\THM{th:Commoner}{Commoner's Theorem}
For any free-choice net $N$ with no isolated places and any marking
	$M_0$ of $N$, the following
two conditions are equivalent:
\begin{enumerate}
\item[a)]
	$(N,M_0)$ is live.
\item[b)]
	Every nonempty siphon $R\subseteq S_N$ 
contains a trap $Q\subseteq R$ such that
$M_0\restr{Q}\neq \mathbf{0}$.
\end{enumerate}
\ENDTHM
\PF
We fix a free-choice net $N$ and establish the two implications.

1. \emph{b)} $\Rightarrow$ \emph{a)}: (This implication holds even if isolated places are present.)\\
We assume that $(N, M_0)$ is nonlive and show that there exists a nonempty siphon $R$ such that every trap $Q \subseteq R$ satisfies $M_0\restr{Q} = \mathbf{0}$. 

Since $M_0$ is nonlive, there exists a DL-marking $M \in \rset{M_0}$ (by Proposition~\ref{prop:dlmarking}). We fix such an $M$ and choose a semi-S-component $X$ of $N$ such that $M\restr{S_X} = \mathbf{0}$ (which exists by Proposition~\ref{prop:MDLtopSCC}).
Observe that $S_X$ is a nonempty siphon, since $S_X \neq \emptyset$
and $\pre{S_X} = T_X \subseteq \post{S_X}$ by
Definition~\ref{def:Snotions}. Furthermore, every trap $Q\subseteq
S_X$ satisfies 
$M\restr{Q}=\mathbf{0}$, which implies $M_0\restr{Q}=\mathbf{0}$
since  $M\in\rset{M_0}$ (recall that $M_0\restr{Q}\neq\mathbf{0}$
implies
$M'\restr{Q}\neq\mathbf{0}$ for every $M'\in\rset{M_0}$).

2. \emph{a)} $\Rightarrow$ \emph{b)}:\\
Now we assume that the fixed free-choice net $N$  has no isolated
places and consider a marking $M_0$
and a nonempty siphon $R$ satisfying $M_0\restr{Q}=\mathbf{0}$ for
the maximal trap $Q\subseteq R$. We will show that $M_0$ is nonlive.
Since $N$ has no isolated places, we have $\pre{R} \subseteq \post{R}
\neq \emptyset$. We complete the proof by constructing a~marking $M \in \rset{M_0}$ at which all transitions in $\post{R}$ are dead.

If the maximal trap $Q\subseteq R$ satisfies $Q = R$, then all
transitions in $\post{R}$ are already dead at $M_0$ (since
$M_0\restr{R} = \mathbf{0}$ and $\pre{R} \subseteq \post{R}$). We thus further
assume that $R \setminus Q$ is nonempty, and consider the
nonempty set $T\leak=\bigcup_{i\in[1,m]}T_i$ of the respective leaking transitions.

Now we stepwise construct an execution Exec from $M_0$ by firing the
leaking transitions in $T\leak$ as frequently as possible, while preserving the
invariant that $Q$ remains unmarked. Having constructed a~prefix $M_0
\gt{\sigma} M$ of Exec (starting with $M_0 \gt{\varepsilon}
M_0$, where $M_0 \restr{Q} = \mathbf{0}$), we prolong it whenever
there exists a~transition in $\post{R}$ that is not dead at $M$. 

In
this case, we select a shortest sequence $M \xrightarrow{\sigma'} M'$
such that $M'$ enables some $t_0 \in \post{R}$. We then extend the
prefix to $M_0 \xrightarrow{\sigma}M\gt{\sigma'} M'$. Note that $M'
\restr{Q} = \mathbf{0}$ because $M \restr{Q} = \mathbf{0}$ by the
invariant, and no transition from
$\post{R}\supseteq\pre{R}\supseteq\pre{Q}$ occurs in $\sigma'$, since an 
occurrence of such a transition in $\sigma'$ would contradict the
minimality of the length of $M \xrightarrow{\sigma'} M'$.
Hence, $\pre{t_0}\cap R\subseteq R\setminus Q\subseteq \pre{T\leak}$ (recall
Proposition~\ref{prop:maxTrapFacts}).
Consequently, there exists some place
$s\in\pre{t_0}\cap\pre{T\leak}$; thus,
$\pre{t_0} = \pre{t}$ for some  $t\in T\leak$,
due to the free-choice property (as illustrated
in Figure~\ref{fig:commoner} for $s=s_{21}$ and $t = t_{21}$).
Hence, $M'$ enables $t$ as well, and we prolong the
prefix of Exec to $M_0\gt{\sigma\sigma'}M'\gt{t}M''$. Since
$M'\restr{Q}=\mathbf{0}$ and
$\post{t}\cap Q=\emptyset$ (by item 3 of Proposition~\ref{prop:maxTrapFacts}), we preserve the
invariant: $M''\restr{Q}=\mathbf{0}$.

We finish the proof by showing that Exec is
finite, ending in a marking $M$ where all transitions in $\post{R}$
are dead. Otherwise, Exec would be an infinite execution that fires
transitions from $T\leak$ infinitely often, while firing no other
transitions from $\post{R}$.
Consider a transition $t\in T\leak$ with the maximal exit index
$\mathit{ei}(t)$ among those fired
infinitely often, and recall the place $s_t$
from Proposition~\ref{prop:maxTrapFacts}.
The marking of $s_t$ is  decreased infinitely often
(whenever $t$ fires) 
but increased only finitely many times
(by leaking transitions with
strictly higher exit indices than $\mathit{ei}(t)$)---a contradiction.
\ENDPF

\bibliographystyle{fundam}
\bibliography{../fc.bib}

\end{document}

%% file: figures/wf_sc.tex


\newboolean{STarcs}
\newboolean{TSarcs}
\newboolean{ShowPlaceLabels}
\newboolean{ShowTransitionLabels}
\newboolean{ShowPlaces}
\newboolean{ShowTransitions}

\pgfkeys{/mysettings/opacity/.initial=1}

\setboolean{STarcs}{true}
\setboolean{TSarcs}{true}
\setboolean{ShowPlaceLabels}{true}
\setboolean{ShowTransitionLabels}{true}
\setboolean{ShowPlaces}{true}
\setboolean{ShowTransitions}{true}

\ifthenelse{\boolean{STarcs}}{
  \def\STopacity{1}
}{
  \def\STopacity{0}
}
\ifthenelse{\boolean{TSarcs}}{
  \def\TSopacity{1}
}{
  \def\TSopacity{0}
}

\ifthenelse{\boolean{ShowPlaceLabels}}{
  \def\PlaceLabelOpacity{1}
}{
  \def\PlaceLabelOpacity{0}
}
\ifthenelse{\boolean{ShowTransitionLabels}}{
  \def\TransitionLabelOpacity{1}
}{
  \def\TransitionLabelOpacity{0}
}
\ifthenelse{\boolean{ShowPlaces}}{
  \def\PlaceOpacity{1}
}{
  \def\PlaceOpacity{0}
}
\ifthenelse{\boolean{ShowTransitions}}{
  \def\TransitionOpacity{1}
}{
  \def\TransitionOpacity{0}
}

\tikzset{
  STarc/.style={-latex, thick, bend left=0, opacity=\STopacity},
  TSarc/.style={-latex, thick, opacity=\TSopacity},
  place/.style={
  	circle,
  	draw,
  	thick,
  	minimum size=0.5cm,
  	opacity=\PlaceOpacity
  },
  transition/.style={
    draw,
    thick,
    minimum size=0.4cm,
    opacity=\TransitionOpacity
  },
  clusterlabel/.style={
    label=below:#1
  },
  clusterbox/.style={
    rectangle,
    color=blue,
    opacity=0.7,
    fill=yellow,
    rounded corners,
    text width=0.5cm
  }
}

\begin{tikzpicture}

\node[place] (s1) at (1,0.5) [label={[opacity=\PlaceLabelOpacity]left:$s_{2}$}] {};
\node[place] (s2) at (2.5,3) [label={[opacity=\PlaceLabelOpacity]left:$s_{3}$}] {};
\node[place] (s3) at (4,3) [label={[opacity=\PlaceLabelOpacity]right:$s_{4}$}] {};
\node[place] (s4) at (1,5.5) [label={[opacity=\PlaceLabelOpacity]left:$s_{1}$}] {};
\node[place] (s5) at (7,4.5) [label={[opacity=\PlaceLabelOpacity]right:$s_{5}$}] {};

\node[transition] (t1) at (2,1.5) [label={[opacity=\TransitionLabelOpacity]left:$t_{3}$}] {};
\node[transition] (t2) at (2,4.5) [label={[opacity=\TransitionLabelOpacity]left:$t_{2}$}] {};
\node[transition] (t3) at (0.5,3) [label={[opacity=\TransitionLabelOpacity]left:$t_{1}$}] {};
\node[transition] (t4) at (7,0.5) [label={[opacity=\TransitionLabelOpacity]right:$t_{4}$}] {};

\fill (2.5,3) circle (2.5pt);
\fill (4,3) circle (2.5pt);

\draw[STarc](s1)edge[](t1);
\draw[STarc](s2)edge[](t2);
\draw[STarc](s3)edge[](t2);
\draw[STarc](s4)edge[](t3);
\draw[STarc, bend right=45](s5)edge[](t4);
\draw[STarc, dashed](s1)edge[](t4);

\draw[TSarc](t1)edge[](s2);
\draw[TSarc](t1)edge[](s3);
\draw[TSarc](t2)edge[](s4);
\draw[TSarc](t3)edge[](s1);
\draw[TSarc, bend right=45](t4)edge[](s5);
\draw[TSarc, dashed](t2)edge[](s5);

\end{tikzpicture}

%% file: figures/five_clusters.tex


\newboolean{STarcs}
\newboolean{TSarcs}
\newboolean{ShowPlaceLabels}
\newboolean{ShowTransitionLabels}
\newboolean{Showtransitions}
\newboolean{Showplaces}

\pgfkeys{/mysettings/opacity/.initial=1}

\setboolean{STarcs}{true}
\setboolean{TSarcs}{true}
\setboolean{ShowPlaceLabels}{true}
\setboolean{ShowTransitionLabels}{true}
\setboolean{Showtransitions}{true}
\setboolean{Showplaces}{true}

\ifthenelse{\boolean{STarcs}}{
  \def\STopacity{1}
}{
  \def\STopacity{0}
}
\ifthenelse{\boolean{TSarcs}}{
  \def\TSopacity{1}
}{
  \def\TSopacity{0}
}

\ifthenelse{\boolean{ShowPlaceLabels}}{
  \def\PlaceLabelOpacity{1}
}{
  \def\PlaceLabelOpacity{0}
}
\ifthenelse{\boolean{ShowTransitionLabels}}{
  \def\TransitionLabelOpacity{1}
}{
  \def\TransitionLabelOpacity{0}
}
\ifthenelse{\boolean{Showtransitions}}{
  \def\PlaceOpacity{1}
}{
  \def\PlaceOpacity{0}
}
\ifthenelse{\boolean{Showplaces}}{
  \def\TransitionOpacity{1}
}{
  \def\TransitionOpacity{0}
}

\tikzset{
  STarc/.style={-latex, thick, bend left=0, opacity=\STopacity},
  TSarc/.style={latex-, thick, opacity=\TSopacity},
  place/.style={
  	circle,
  	draw,
  	thick,
  	minimum size=0.5cm,
  	opacity=\PlaceOpacity
  },
  transition/.style={
    draw,
    thick,
    minimum size=0.4cm,
    opacity=\TransitionOpacity
  },
  clusterlabel/.style={
    label=below:#1
  },
  clusterbox/.style={
    rectangle,
    color=blue,
    opacity=0.7,
    fill=yellow,
    rounded corners,
    text width=0.5cm
  }
}


\begin{tikzpicture}[yscale=0.8,xscale=0.85]

\node[clusterbox, minimum height=1.6cm, minimum width=2.6cm] (C1) at (2.25,6.5) {};
\node[clusterlabel] at (4,6.9) {C1};

\node[place] (s11) at (3,6) [label={[opacity=\PlaceLabelOpacity]-100:$s_{11}$}] {};
\node[place] (s12) at (3,7) [label={[opacity=\PlaceLabelOpacity]above:$s_{12}$}] {};

\node[transition] (t11) at (1.5,6) [label={[opacity=\TransitionLabelOpacity]below:$t_{11}$}] {};
\node[transition] (t12) at (1.5,7) [label={[opacity=\TransitionLabelOpacity]above:$t_{12}$}] {};

\draw[STarc](s11)edge[](t11);
\draw[STarc](s11)edge[](t12);
\draw[STarc](s12)edge[](t11);
\draw[STarc](s12)edge[](t12);

\node[clusterbox, minimum height=2.5cm, minimum width=2.3cm, rotate=45] (C2) at (1.5,1.5) {};
\node[clusterlabel] at (0,2.5) {C2};

\node[place] (s21) at (1.5,2.5) [label={[opacity=\PlaceLabelOpacity]0:$s_{21}$}] {};
\node[place] (s22) at (0.5,1.5) [label={[opacity=\PlaceLabelOpacity]left:$s_{22}$}] {};

\node[transition, rotate=45] (t21) at (2.5,1.5) [label={[opacity=\TransitionLabelOpacity]below:$t_{21}$}] {};
\node[transition, rotate=45] (t22) at (1.5,0.5) [label={[opacity=\TransitionLabelOpacity]below left:$t_{22}$}] {};

\draw[STarc](s21)edge[](t21);
\draw[STarc](s21)edge[](t22);
\draw[STarc](s22)edge[](t21);
\draw[STarc](s22)edge[](t22);

\node[clusterbox, minimum height=2cm, minimum width=3.2cm] (C3) at (6.3,3.6) {};
\node[clusterlabel] at (7.3,5.1) {C3};

\node[place] (s31) at (5,3) [label={[opacity=\PlaceLabelOpacity]left:$s_{31}$}] {};
\node[place, line width=0.06cm] (s32) at (6.5,3) [label={[opacity=\PlaceLabelOpacity]right:$s_{32}$}] {};

\node[transition] (t31) at (5,4.3) [label={[opacity=\TransitionLabelOpacity, inner sep=0.1mm, label distance=0.2mm]220:$t_{31}$}] {};
\node[transition] (t32) at (6.5,4.3) [label={[opacity=\TransitionLabelOpacity, inner sep=0.1mm, label distance=0.2mm]170:$t_{32}$}] {};
\node[transition, line width=0.06cm] (t33) at (7.5,4.3) [label={[opacity=\TransitionLabelOpacity]right:$t_{33}$}] {};

\draw[STarc](s31)edge[](t31);
\draw[STarc](s31)edge[](t32);
\draw[STarc](s32)edge[](t31);
\draw[STarc](s32)edge[](t32);
\draw[STarc](s31)edge[](t33);
\draw[STarc, line width=0.06cm](s32)edge[](t33);

\node[clusterbox, minimum height=2cm, minimum width=3.1cm] (C4) at (9.2,0.7) {};
\node[clusterlabel] at (11.3,0.8) {C4};

\node[place] (s41) at (8,0) [label={[opacity=\PlaceLabelOpacity]left:$s_{41}$}] {};
\node[place] (s42) at (9,0) [label={[opacity=\PlaceLabelOpacity]-50:$s_{42}$}] {};
\node[place, line width=0.06cm] (s43) at (10,0) [label={[opacity=\PlaceLabelOpacity]right:$s_{43}$}] {};

\node[transition] (t41) at (9,1.5) [label={[opacity=\TransitionLabelOpacity]120:$t_{41}$}] {};
\node[transition, line width=0.06cm] (t42) at (10,1.5) [label={[opacity=\TransitionLabelOpacity]0:$t_{42}$}] {};

\draw[STarc](s41)edge[](t41);
\draw[STarc](s41)edge[](t42);
\draw[STarc](s42)edge[](t41);
\draw[STarc](s42)edge[](t42);
\draw[STarc](s43)edge[](t41);
\draw[STarc, line width=0.06cm](s43)edge[](t42);

\begin{scope}[shift={(-3,4)}]
\node[clusterbox, minimum height=1.6cm, minimum width=2.3cm] (C5) at (12.7,2.5) {};
\node[clusterlabel] at (11.7,3.7) {C5};

\node[place, line width=0.06cm] (s51) at (12,2) [label={[opacity=\PlaceLabelOpacity]left:$s_{51}$}] {};
\node[place, line width=0.06cm] (s52) at (13,2) [label={[opacity=\PlaceLabelOpacity]right:$s_{52}$}] {};

\node[transition, line width=0.06cm] (t51) at (12,3) [label={[opacity=\TransitionLabelOpacity]left:$t_{51}$}] {};

\draw[STarc, line width=0.06cm](s51)edge[](t51);
\draw[STarc, line width=0.06cm](s52)edge[](t51);
\end{scope}


\draw[TSarc](s12.north east)to[out=20,in=160,looseness=4](t12.north west);
\draw [TSarc](s41.south west) 
 .. controls ++(15:-1.5) and ++(170:-1.5) .. (0,-1)
 .. controls ++(170:1.5) and ++(170:1.5) .. (t12.west);
\draw[TSarc, bend left=40](s21)edge[](t11);

\draw[TSarc, bend left=45](s12)edge[](t21);
\draw[TSarc, bend left=40](s31)edge[](t21);
\draw[TSarc](s42) .. controls +(0,-2) and +(0.5,-2) ..(t22);
\draw[TSarc, bend left=25](s11)to[](t21);

\draw[TSarc, bend left=55](s22)edge[](t31);
\draw[TSarc, bend left=30](s11)edge[](t31);
\draw[TSarc, bend left=45](s11)edge[](t32);
\draw[TSarc, bend left=10, line width=0.06cm](s51)edge[](t33);

\draw[TSarc](s41) .. controls +(-2,1) and +(-4,0.5) .. (t41);
\draw[TSarc, line width=0.06cm](s32)to[out=300,in=120](t42);
\draw[TSarc, line width=0.06cm](s52) .. controls +(1,-2) and +(1,1) .. (t42);

\draw[TSarc, line width=0.06cm](s43) .. controls +(3,-4) and +(4,4) .. (t51);

\end{tikzpicture}

%% file: figures/bott_scc_alloc.tex



\newboolean{STarcs}
\newboolean{TSarcs}
\newboolean{ShowPlaceLabels}
\newboolean{ShowTransitionLabels}
\newboolean{ShowPlaces}
\newboolean{ShowTransitions}

\pgfkeys{/mysettings/opacity/.initial=1}

\setboolean{STarcs}{true}
\setboolean{TSarcs}{true}
\setboolean{ShowPlaceLabels}{true}
\setboolean{ShowTransitionLabels}{true}
\setboolean{ShowPlaces}{true}
\setboolean{ShowTransitions}{true}

\ifthenelse{\boolean{STarcs}}{
  \def\STopacity{1}
}{
  \def\STopacity{0}
}
\ifthenelse{\boolean{TSarcs}}{
  \def\TSopacity{1}
}{
  \def\TSopacity{0}
}

\ifthenelse{\boolean{ShowPlaceLabels}}{
  \def\PlaceLabelOpacity{1}
}{
  \def\PlaceLabelOpacity{0}
}
\ifthenelse{\boolean{ShowTransitionLabels}}{
  \def\TransitionLabelOpacity{1}
}{
  \def\TransitionLabelOpacity{0}
}
\ifthenelse{\boolean{ShowPlaces}}{
  \def\PlaceOpacity{1}
}{
  \def\PlaceOpacity{0}
}
\ifthenelse{\boolean{ShowTransitions}}{
  \def\TransitionOpacity{1}
}{
  \def\TransitionOpacity{0}
}

\tikzset{
  STarc/.style={-latex, thick, bend left=0, opacity=\STopacity},
  TSarc/.style={latex-, thick, opacity=\TSopacity},
  place/.style={
  	circle,
  	draw,
  	thick,
  	minimum size=0.5cm,
  	opacity=\PlaceOpacity
  },
  transition/.style={
    draw,
    thick,
    minimum size=0.4cm,
    opacity=\TransitionOpacity
  },
  clusterlabel/.style={
    label=below:#1
  },
  clusterbox/.style={
    rectangle,
    color=blue,
    opacity=0.7,
    fill=yellow,
    rounded corners,
    text width=0.5cm
  }
}


\begin{tikzpicture}[yscale=0.8,xscale=0.85]

\node[clusterbox, minimum height=1.6cm, minimum width=2.6cm] (C1) at (2.25,6.5) {};
\node[clusterlabel] at (4,6.9) {C1};

\node[place, line width=0.06cm] (s11) at (3,6) [label={[opacity=\PlaceLabelOpacity]-100:$s_{11}$}] {};
\node[place, line width=0.06cm] (s12) at (3,7) [label={[opacity=\PlaceLabelOpacity]above:$s_{12}$}] {};

\node[transition, line width=0.06cm] (t11) at (1.5,6) [label={[opacity=\TransitionLabelOpacity]below:$t_{11}$}] {};

\draw[STarc, line width=0.06cm](s11)edge[](t11);
\draw[STarc, line width=0.06cm](s12)edge[](t11);

\node[clusterbox, minimum height=2.5cm, minimum width=2.3cm, rotate=45] (C2) at (1.5,1.5) {};
\node[clusterlabel] at (0,2.5) {C2};

\node[place, line width=0.06cm] (s21) at (1.5,2.5) [label={[opacity=\PlaceLabelOpacity]0:$s_{21}$}] {};
\node[place, line width=0.06cm] (s22) at (0.5,1.5) [label={[opacity=\PlaceLabelOpacity]left:$s_{22}$}] {};

\node[transition, rotate=45, line width=0.06cm] (t21) at (2.5,1.5) [label={[opacity=\TransitionLabelOpacity]below:$t_{21}$}] {};

\draw[STarc, line width=0.06cm](s21)edge[](t21);
\draw[STarc, line width=0.06cm](s22)edge[](t21);

\node[clusterbox, minimum height=2cm, minimum width=3.2cm] (C3) at (6.3,3.6) {};
\node[clusterlabel] at (7.3,5.1) {C3};

\node[place, line width=0.06cm] (s31) at (5,3) [label={[opacity=\PlaceLabelOpacity]left:$s_{31}$}] {};
\node[place] (s32) at (6.5,3) [label={[opacity=\PlaceLabelOpacity]right:$s_{32}$}] {};

\node[transition, line width=0.06cm] (t31) at (5,4.3) [label={[opacity=\TransitionLabelOpacity, inner sep=0.1mm, label distance=0.2mm]220:$t_{31}$}] {};

\draw[STarc, line width=0.06cm](s31)edge[](t31);
\draw[STarc, red, line width=0.06cm,
	line join=round,thick,decorate,
decoration={zigzag,segment length=1mm,amplitude=0.2mm,post=lineto,post
	length=4pt}](s32)to[](t31);

\node[clusterbox, minimum height=2cm, minimum width=3.1cm] (C4) at (9.2,0.7) {};
\node[clusterlabel] at (11.3,0.8) {C4};

\node[place] (s41) at (8,0) [label={[opacity=\PlaceLabelOpacity]left:$s_{41}$}] {};
\node[place] (s42) at (9,0) [label={[opacity=\PlaceLabelOpacity]-50:$s_{42}$}] {};
\node[place] (s43) at (10,0) [label={[opacity=\PlaceLabelOpacity]right:$s_{43}$}] {};

\node[transition] (t42) at (10,1.5) [label={[opacity=\TransitionLabelOpacity]0:$t_{42}$}] {};

\draw[STarc](s41)edge[](t42);
\draw[STarc](s42)edge[](t42);
\draw[STarc](s43)edge[](t42);

\begin{scope}[shift={(-3,4)}]
\node[clusterbox, minimum height=1.6cm, minimum width=2.3cm] (C5) at (12.7,2.5) {};
\node[clusterlabel] at (11.7,3.7) {C5};

\node[place] (s51) at (12,2) [label={[opacity=\PlaceLabelOpacity]left:$s_{51}$}] {};
\node[place] (s52) at (13,2) [label={[opacity=\PlaceLabelOpacity]right:$s_{52}$}] {};

\node[transition] (t51) at (12,3) [label={[opacity=\TransitionLabelOpacity]left:$t_{51}$}] {};

\draw[STarc](s51)edge[](t51);
\draw[STarc](s52)edge[](t51);
\end{scope}


\draw[TSarc, bend left=40, line width=0.06cm](s21)edge[](t11);

\draw[TSarc, bend left=45, line width=0.06cm](s12)edge[](t21);
\draw[TSarc, bend left=40, line width=0.06cm](s31)edge[](t21);
\draw[TSarc, bend left=25, red, line join=round,
      line width=0.06cm,
      decorate,
      decoration={zigzag, segment length=1mm, amplitude=0.25mm, pre length=2mm, post length=2mm}]
      (s11) to (t21);

\draw[TSarc, bend left=55, line width=0.06cm](s22)edge[](t31);
\draw[TSarc, bend left=30, line width=0.06cm](s11)edge[](t31);

\draw[TSarc](s32)to[out=300,in=120](t42);
\draw[TSarc](s52) .. controls +(1,-2) and +(1,1) .. (t42);

\draw[TSarc](s43) .. controls +(3,-4) and +(4,4) .. (t51);

\end{tikzpicture}

%% file: figures/t_covered.tex


\newboolean{STarcs}
\newboolean{TSarcs}
\newboolean{ShowPlaceLabels}
\newboolean{ShowTransitionLabels}
\newboolean{Showtransitions}
\newboolean{Showplaces}

\pgfkeys{/mysettings/opacity/.initial=1}

\setboolean{STarcs}{true}
\setboolean{TSarcs}{true}
\setboolean{ShowPlaceLabels}{true}
\setboolean{ShowTransitionLabels}{true}
\setboolean{Showtransitions}{true}
\setboolean{Showplaces}{true}

\ifthenelse{\boolean{STarcs}}{
  \def\STopacity{1}
}{
  \def\STopacity{0}
}
\ifthenelse{\boolean{TSarcs}}{
  \def\TSopacity{1}
}{
  \def\TSopacity{0}
}

\ifthenelse{\boolean{ShowPlaceLabels}}{
  \def\PlaceLabelOpacity{1}
}{
  \def\PlaceLabelOpacity{0}
}
\ifthenelse{\boolean{ShowTransitionLabels}}{
  \def\TransitionLabelOpacity{1}
}{
  \def\TransitionLabelOpacity{0}
}
\ifthenelse{\boolean{Showtransitions}}{
  \def\PlaceOpacity{1}
}{
  \def\PlaceOpacity{0}
}
\ifthenelse{\boolean{Showplaces}}{
  \def\TransitionOpacity{1}
}{
  \def\TransitionOpacity{0}
}

\tikzset{
  STarc/.style={-latex, thick, bend left=0, opacity=\STopacity},
  TSarc/.style={-latex, thick, opacity=\TSopacity},
  place/.style={
  	circle,
  	draw,
  	thick,
  	minimum size=0.5cm,
  	opacity=\PlaceOpacity
  },
  transition/.style={
    draw,
    thick,
    minimum size=0.4cm,
    opacity=\TransitionOpacity
  },
  clusterlabel/.style={
    label=below:#1
  },
  clusterbox/.style={
    rectangle,
    color=blue,
    opacity=0.7,
    fill=yellow,
    rounded corners,
    text width=0.5cm
  }
}

\begin{tikzpicture}[rotate=90, xshift=-1.2cm, yshift=2.0cm]

  \node[clusterbox, minimum height=1.3cm, minimum width=1.8cm] (C1) at (1,6.5) {};

  \node[place, line width=0.06cm] (s1) at (1,7) [label=above:$s_1$] {};
  \node[transition, line width=0.06cm] (t1) at (1,6) [label=above:$t_1$] {};

  \node[clusterbox, minimum height=1.6cm, minimum width=2cm] (C3) at (2,4.2) {};

  \node[place, line width=0.06cm] (s2) at (2,4.8) [label=left:$s_2$] {};
  \node[transition, line width=0.06cm] (t3) at (1.5,3.6) [label=190:$t_{3}$] {};
  \node[transition] (t2) at (2.5,3.6) [label=above:$t_2$] {};

  \node[clusterbox, minimum height=1.6cm, minimum width=2cm] (C2) at (0,4.2) {};

  \node[place, line width=0.06cm] (s3) at (0,4.8) [label=left:$s_3$] {};
  \node[transition, line width=0.06cm] (t5) at (-0.5,3.6) [label=below:$t_5$] {};
  \node[transition] (t4) at (0.5,3.6) [label=175:$t_{4}$] {};

  \node[clusterbox, minimum height=1.6cm, minimum width=2cm] (C5) at (2,1.8) {};

  \node[place] (s5) at (1.5,2.4) [label=right:$s_{5}$] {};
  \node[place] (s4) at (2.5,2.4) [label=right:$s_4$] {};
  \node[transition] (t6) at (2,1.2) [label=right:$t_6$] {};

  \node[clusterbox, minimum height=1.6cm, minimum width=2cm] (C4) at (0,1.8) {};

  \node[place, line width=0.06cm] (s7) at (-0.5,2.4) [label=right:$s_7$] {};
  \node[place, line width=0.06cm] (s6) at (0.5,2.4) [label=right:$s_{6}$] {};
  \node[transition, line width=0.06cm] (t7) at (0,1.2) [label=right:$t_7$] {};

  \draw[STarc, line width=0.06cm] (s1) -- (t1);
  \draw[TSarc, line width=0.06cm] (t1) -- (s2);
  \draw[TSarc, line width=0.06cm] (t1) -- (s3);
  \draw[STarc] (s2) -- (t2);
  \draw[TSarc, line width=0.06cm] (s2) -- (t3);
  \draw[STarc] (s3) -- (t4);
  \draw[STarc, line width=0.06cm] (s3) -- (t5);

  \draw[TSarc] (t2) -- (s4);
  \draw[TSarc, line width=0.06cm] (t3) -- (s6);
  \draw[TSarc] (t4) -- (s5);
  \draw[TSarc, line width=0.06cm] (t5) -- (s7);

  \draw[STarc] (s4) -- (t6);
  \draw[STarc, line width=0.06cm] (s6) -- (t7);
  \draw[STarc] (s5) -- (t6);
  \draw[STarc, line width=0.06cm] (s7) -- (t7);

  \draw[TSarc, looseness=1.3] (t6) edge[out=316, in=40] (s1);
  \draw[TSarc, looseness=1.3, line width=0.06cm] (t7) edge[out=230, in=150] (s1);

\end{tikzpicture}

%% file: figures/proper_covered.tex


\newboolean{STarcs}
\newboolean{TSarcs}
\newboolean{ShowPlaceLabels}
\newboolean{ShowTransitionLabels}
\newboolean{Showtransitions}
\newboolean{Showplaces}

\pgfkeys{/mysettings/opacity/.initial=1}

\setboolean{STarcs}{true}
\setboolean{TSarcs}{true}
\setboolean{ShowPlaceLabels}{true}
\setboolean{ShowTransitionLabels}{true}
\setboolean{Showtransitions}{true}
\setboolean{Showplaces}{true}

\ifthenelse{\boolean{STarcs}}{
  \def\STopacity{1}
}{
  \def\STopacity{0}
}
\ifthenelse{\boolean{TSarcs}}{
  \def\TSopacity{1}
}{
  \def\TSopacity{0}
}

\ifthenelse{\boolean{ShowPlaceLabels}}{
  \def\PlaceLabelOpacity{1}
}{
  \def\PlaceLabelOpacity{0}
}
\ifthenelse{\boolean{ShowTransitionLabels}}{
  \def\TransitionLabelOpacity{1}
}{
  \def\TransitionLabelOpacity{0}
}
\ifthenelse{\boolean{Showtransitions}}{
  \def\PlaceOpacity{1}
}{
  \def\PlaceOpacity{0}
}
\ifthenelse{\boolean{Showplaces}}{
  \def\TransitionOpacity{1}
}{
  \def\TransitionOpacity{0}
}

\tikzset{
  STarc/.style={-latex, thick, bend left=0, opacity=\STopacity},
  TSarc/.style={-latex, thick, opacity=\TSopacity},
  place/.style={
  	circle,
  	draw,
  	thick,
  	minimum size=0.5cm,
  	opacity=\PlaceOpacity
  },
  transition/.style={
    draw,
    thick,
    minimum size=0.4cm,
    opacity=\TransitionOpacity
  },
  clusterlabel/.style={
    label=below:#1
  },
  clusterbox/.style={
    rectangle,
    color=blue,
    opacity=0.7,
    fill=yellow,
    rounded corners,
    text width=0.5cm
  }
}

\begin{tikzpicture}[rotate=90, xshift=-1.2cm, yshift=2.0cm]

  \node[clusterbox, minimum height=1.3cm, minimum width=1.8cm] (C1) at (1,6.5) {};

  \node[place, line width=0.06cm] (s1) at (1,7) [label=above:$s_1$] {};
  \node[transition, line width=0.06cm] (t1) at (1,6) [label=above:$t_1$] {};

  \node[clusterbox, minimum height=1.6cm, minimum width=2cm] (C3) at (2,4.2) {};

  \node[place, line width=0.06cm] (s2) at (2,4.8) [label=left:$s_2$] {};
  \node[transition] (t3) at (1.5,3.6) [label=190:$t_{3}$] {};
  \node[transition, line width=0.06cm] (t2) at (2.5,3.6) [label=above:$t_2$] {};

  \node[clusterbox, minimum height=1.6cm, minimum width=2cm] (C2) at (0,4.2) {};

  \node[place, line width=0.06cm] (s3) at (0,4.8) [label=left:$s_3$] {};
  \node[transition, line width=0.06cm] (t5) at (-0.5,3.6) [label=below:$t_5$] {};
  \node[transition] (t4) at (0.5,3.6) [label=175:$t_{4}$] {};

  \node[clusterbox, minimum height=1.6cm, minimum width=2cm] (C5) at (2,1.8) {};

  \node[place] (s5) at (1.5,2.4) [label=right:$s_{5}$] {};
  \node[place, line width=0.06cm] (s4) at (2.5,2.4) [label=right:$s_4$] {};
  \node[transition, line width=0.06cm] (t6) at (2,1.2) [label=right:$t_6$] {};

  \node[clusterbox, minimum height=1.6cm, minimum width=2cm] (C4) at (0,1.8) {};

  \node[place, line width=0.06cm] (s7) at (-0.5,2.4) [label=right:$s_7$] {};
  \node[place] (s6) at (0.5,2.4) [label=right:$s_{6}$] {};
  \node[transition, line width=0.06cm] (t7) at (0,1.2) [label=right:$t_7$] {};

  \draw[STarc, line width=0.06cm] (s1) -- (t1);
  \draw[TSarc, line width=0.06cm] (t1) -- (s2);
  \draw[TSarc, line width=0.06cm] (t1) -- (s3);
  \draw[STarc, line width=0.06cm] (s2) -- (t2);
  \draw[TSarc] (s2) -- (t3);
  \draw[STarc] (s3) -- (t4);
  \draw[STarc, line width=0.06cm] (s3) -- (t5);

  \draw[TSarc, line width=0.06cm] (t2) -- (s4);
  \draw[TSarc] (t3) -- (s6);
  \draw[TSarc] (t4) -- (s5);
  \draw[TSarc, line width=0.06cm] (t5) -- (s7);

  \draw[STarc, line width=0.06cm] (s4) -- (t6);
  \draw[STarc] (s6) -- (t7);
  \draw[STarc] (s5) -- (t6);
  \draw[STarc, line width=0.06cm] (s7) -- (t7);

  \draw[TSarc, looseness=1.3, line width=0.06cm] (t6) edge[out=316, in=40] (s1);
  \draw[TSarc, looseness=1.3, line width=0.06cm] (t7) edge[out=230, in=150] (s1);

\end{tikzpicture}

%% file: figures/five_clusters_dual.tex

\definecolor{pnblue}{RGB}{109, 150, 242}

\newboolean{STarcs}
\newboolean{TSarcs}
\newboolean{ShowPlaceLabels}
\newboolean{ShowTransitionLabels}
\newboolean{Showtransitions}
\newboolean{Showplaces}

\pgfkeys{/mysettings/opacity/.initial=1}

\setboolean{STarcs}{true}
\setboolean{TSarcs}{true}
\setboolean{ShowPlaceLabels}{true}
\setboolean{ShowTransitionLabels}{true}
\setboolean{Showtransitions}{true}
\setboolean{Showplaces}{true}

\ifthenelse{\boolean{STarcs}}{
  \def\STopacity{1}
}{
  \def\STopacity{0}
}
\ifthenelse{\boolean{TSarcs}}{
  \def\TSopacity{1}
}{
  \def\TSopacity{0}
}

\ifthenelse{\boolean{ShowPlaceLabels}}{
  \def\PlaceLabelOpacity{1}
}{
  \def\PlaceLabelOpacity{0}
}
\ifthenelse{\boolean{ShowTransitionLabels}}{
  \def\TransitionLabelOpacity{1}
}{
  \def\TransitionLabelOpacity{0}
}
\ifthenelse{\boolean{Showtransitions}}{
  \def\PlaceOpacity{1}
}{
  \def\PlaceOpacity{0}
}
\ifthenelse{\boolean{Showplaces}}{
  \def\TransitionOpacity{1}
}{
  \def\TransitionOpacity{0}
}

\tikzset{
  STarc/.style={-latex, thick, bend left=0, opacity=\STopacity},
  TSarc/.style={-latex, thick, opacity=\TSopacity},
  place/.style={
  	circle,
  	draw,
  	thick,
  	minimum size=0.5cm,
  	opacity=\PlaceOpacity
  },
  transition/.style={
    draw,
    thick,
    minimum size=0.4cm,
    opacity=\TransitionOpacity
  },
  clusterlabel/.style={
    label=below:#1
  },
  clusterbox/.style={
    rectangle,
    color=pnblue,
    opacity=0.7,
    fill=yellow,
    rounded corners,
    text width=0.5cm
  }
}

\begin{tikzpicture}[yscale=0.8,xscale=0.85] 

\node[clusterbox, minimum height=1.6cm, minimum width=2.6cm] (C1) at (2.25,6.5) {};
\node[clusterlabel] at (4,6.9) {C1};

\node[transition, fill=pnblue, line width=0.06cm] (t11) at (3,6) [label={[opacity=\TransitionLabelOpacity]-100:$t_{11}$}] {};
\node[transition, fill=pnblue, line width=0.06cm] (t12) at (3,7) [label={[opacity=\TransitionLabelOpacity]above:$t_{12}$}] {};

\node[place, fill=pnblue, line width=0.06cm] (s11) at (1.5,6) [label={[opacity=\PlaceLabelOpacity]below:$s_{11}$}] {};
\node[place] (s12) at (1.5,7) [label={[opacity=\PlaceLabelOpacity]above:$s_{12}$}] {};

\draw[STarc, line width=0.06cm](s11)edge[](t11);
\draw[STarc, line width=0.06cm](s11)edge[](t12);
\draw[STarc](s12)edge[](t11);
\draw[STarc](s12)edge[](t12);

\node[clusterbox, minimum height=2.5cm, minimum width=2.3cm, rotate=45] (C2) at (1.5,1.5) {};
\node[clusterlabel] at (0,2.5) {C2};

\node[transition, rotate=45, fill=pnblue, line width=0.06cm] (t21) at (1.5,2.5) [label={[opacity=\TransitionLabelOpacity]0:$t_{21}$}] {};
\node[transition, rotate=45, fill=pnblue, line width=0.06cm] (t22) at (0.5,1.5) [label={[opacity=\TransitionLabelOpacity]left:$t_{22}$}] {};

\node[place, fill=pnblue, line width=0.06cm] (s21) at (2.5,1.5) [label={[opacity=\PlaceLabelOpacity]below:$s_{21}$}] {};
\node[place] (s22) at (1.5,0.5) [label={[opacity=\PlaceLabelOpacity]below left:$s_{22}$}] {};

\draw[STarc, line width=0.06cm](s21)edge[](t21);
\draw[STarc, line width=0.06cm](s21)edge[](t22);
\draw[STarc](s22)edge[](t21);
\draw[STarc](s22)edge[](t22);

\node[clusterbox, minimum height=2cm, minimum width=3.2cm] (C3) at (6.3,3.6) {};
\node[clusterlabel] at (7.3,5.1) {C3};

\node[transition, fill=pnblue, line width=0.06cm] (t31) at (5,3) [label={[opacity=\TransitionLabelOpacity]left:$t_{31}$}] {};
\node[transition, fill=pnblue] (t32) at (6.5,3) [label={[opacity=\TransitionLabelOpacity]right:$t_{32}$}] {};

\node[place, fill=pnblue, line width=0.06cm] (s31) at (5,4.3) [label={[opacity=\PlaceLabelOpacity, inner sep=0.1mm, label distance=0.2mm]220:$s_{31}$}] {};
\node[place] (s32) at (6.5,4.3) [label={[opacity=\PlaceLabelOpacity, inner sep=0.1mm, label distance=0.2mm]170:$s_{32}$}] {};
\node[place] (s33) at (7.5,4.3) [label={[opacity=\PlaceLabelOpacity]right:$s_{33}$}] {};

\draw[STarc, line width=0.06cm](s31)edge[](t31);
\draw[STarc, red, 
	line join=round,thick,decorate,
decoration={zigzag,segment length=1mm,amplitude=0.2mm,post=lineto,post length=4pt}](s31)to[](t32);
\draw[STarc](s32)edge[](t31);
\draw[STarc](s32)edge[](t32);
\draw[STarc](s33)edge[](t31);
\draw[STarc](s33)edge[](t32);

\node[clusterbox, minimum height=2cm, minimum width=3.1cm] (C4) at (9.2,0.7) {};
\node[clusterlabel] at (11.3,0.8) {C4};

\node[transition, fill=pnblue] (t41) at (8,0) [label={[opacity=\TransitionLabelOpacity]left:$t_{41}$}] {};
\node[transition, fill=pnblue] (t42) at (9,0) [label={[opacity=\TransitionLabelOpacity]-50:$t_{42}$}] {};
\node[transition, fill=pnblue] (t43) at (10,0) [label={[opacity=\TransitionLabelOpacity]right:$t_{43}$}] {};

\node[place] (s41) at (9,1.5) [label={[opacity=\PlaceLabelOpacity]120:$s_{41}$}] {};
\node[place, fill=pnblue] (s42) at (10,1.5) [label={[opacity=\PlaceLabelOpacity]0:$s_{42}$}] {};

\draw[STarc](s41)edge[](t41);
\draw[STarc](s41)edge[](t42);
\draw[STarc](s41)edge[](t43);
\draw[STarc](s42)edge[](t41);
\draw[STarc](s42)edge[](t42);
\draw[STarc](s42)edge[](t43);

\begin{scope}[shift={(-3,4)}]
\node[clusterbox, minimum height=1.6cm, minimum width=2.3cm] (C5) at (12.7,2.5) {};
\node[clusterlabel] at (11.7,3.7) {C5};

\node[transition, fill=pnblue] (t51) at (12,2) [label={[opacity=\TransitionLabelOpacity]left:$t_{51}$}] {};
\node[transition, fill=pnblue] (t52) at (13,2) [label={[opacity=\TransitionLabelOpacity]right:$t_{52}$}] {};

\node[place, fill=pnblue] (s51) at (12,3) [label={[opacity=\PlaceLabelOpacity]left:$s_{51}$}] {};

\draw[STarc](s51)edge[](t51);
\draw[STarc](s51)edge[](t52);
\end{scope}


\draw[TSarc](t12.north east)to[out=20,in=160,looseness=4](s12.north west);
\draw [TSarc](t41.south west) 
 .. controls ++(15:-1.5) and ++(170:-1.5) .. (0,-1)
 .. controls ++(170:1.5) and ++(170:1.5) .. (s12.west);
\draw[TSarc, bend left=40, line width=0.06cm](t21)edge[](s11);

\draw[TSarc, bend left=45, line width=0.06cm](t12)edge[](s21);
\draw[TSarc, bend left=40, line width=0.06cm](t31)edge[](s21);
\draw[TSarc](t42) .. controls +(0,-2) and +(0.5,-2) ..(s22);
\draw[TSarc, bend left=25, red, line join=round,
      line width=0.06cm,
      decorate,
      decoration={zigzag, segment length=1mm, amplitude=0.25mm, pre length=2mm, post length=2mm}]
      (t11) to (s21);

\draw[TSarc, bend left=55, line width=0.06cm](t22)edge[](s31);
\draw[TSarc, bend left=30, line width=0.06cm](t11)edge[](s31);
\draw[TSarc, bend left=45](t11)edge[](s32);
\draw[TSarc, bend left=10](t51)edge[](s33);

\draw[TSarc](t41) .. controls +(-2,1) and +(-4,0.5) .. (s41);
\draw[TSarc](t32)to[out=300,in=120](s42);
\draw[TSarc](t52) .. controls +(1,-2) and +(1,1) .. (s42);

\draw[TSarc](t43) .. controls +(3,-4) and +(4,4) .. (s51);

\end{tikzpicture}

%% file: figures/dl_marking_dual.tex

\definecolor{pnblue}{RGB}{109, 150, 242}

\newboolean{STarcs}
\newboolean{TSarcs}
\newboolean{ShowPlaceLabels}
\newboolean{ShowTransitionLabels}
\newboolean{Showtransitions}
\newboolean{Showplaces}

\pgfkeys{/mysettings/opacity/.initial=1}

\setboolean{STarcs}{true}
\setboolean{TSarcs}{true}
\setboolean{ShowPlaceLabels}{true}
\setboolean{ShowTransitionLabels}{true}
\setboolean{Showtransitions}{true}
\setboolean{Showplaces}{true}

\ifthenelse{\boolean{STarcs}}{
  \def\STopacity{1}
}{
  \def\STopacity{0}
}
\ifthenelse{\boolean{TSarcs}}{
  \def\TSopacity{1}
}{
  \def\TSopacity{0}
}

\ifthenelse{\boolean{ShowPlaceLabels}}{
  \def\PlaceLabelOpacity{1}
}{
  \def\PlaceLabelOpacity{0}
}
\ifthenelse{\boolean{ShowTransitionLabels}}{
  \def\TransitionLabelOpacity{1}
}{
  \def\TransitionLabelOpacity{0}
}
\ifthenelse{\boolean{Showtransitions}}{
  \def\PlaceOpacity{1}
}{
  \def\PlaceOpacity{0}
}
\ifthenelse{\boolean{Showplaces}}{
  \def\TransitionOpacity{1}
}{
  \def\TransitionOpacity{0}
}

\tikzset{
  STarc/.style={-latex, thick, bend left=0, opacity=\STopacity},
  TSarc/.style={-latex, thick, opacity=\TSopacity},
  place/.style={
  	circle,
  	draw,
  	thick,
  	minimum size=0.5cm,
  	opacity=\PlaceOpacity
  },
  transition/.style={
    draw,
    thick,
    minimum size=0.4cm,
    opacity=\TransitionOpacity
  },
  clusterlabel/.style={
    label=below:#1
  },
  clusterbox/.style={
    rectangle,
    color=pnblue,
    opacity=0.7,
    fill=yellow,
    rounded corners,
    text width=0.5cm
  }
}

\begin{tikzpicture}[yscale=0.8,xscale=0.85]

\node[clusterbox, minimum height=1.6cm, minimum width=2.6cm] (C1) at (2.25,6.5) {};
\node[clusterlabel] at (4,6.9) {C1};

\node[transition, fill=pnblue, line width=0.06cm] (t11) at (3,6) [label={[opacity=\TransitionLabelOpacity]-100:$t_{11}$}] {};
\node[transition, fill=pnblue, line width=0.06cm] (t12) at (3,7) [label={[opacity=\TransitionLabelOpacity]above:$t_{12}$}] {};

\node[place, fill=pnblue, line width=0.06cm] (s11) at (1.5,6) [label={[opacity=\PlaceLabelOpacity]below:$s_{11}$}] {};
\node[place] (s12) at (1.5,7) [label={[opacity=\PlaceLabelOpacity]above:$s_{12}$}] {};

\fill (1.5,7) circle (2.5pt);

\draw[STarc, line width=0.06cm](s11)edge[](t11);
\draw[STarc, line width=0.06cm](s11)edge[](t12);
\draw[STarc](s12)edge[](t11);
\draw[STarc](s12)edge[](t12);

\node[clusterbox, minimum height=2.5cm, minimum width=2.3cm, rotate=45] (C2) at (1.5,1.5) {};
\node[clusterlabel] at (0,2.5) {C2};

\node[transition, rotate=45, fill=pnblue, line width=0.06cm] (t21) at (1.5,2.5) [label={[opacity=\TransitionLabelOpacity]0:$t_{21}$}] {};
\node[transition, rotate=45, fill=pnblue, line width=0.06cm] (t22) at (0.5,1.5) [label={[opacity=\TransitionLabelOpacity]left:$t_{22}$}] {};

\node[place, fill=pnblue, line width=0.06cm] (s21) at (2.5,1.5) [label={[opacity=\PlaceLabelOpacity]below:$s_{21}$}] {};
\node[place] (s22) at (1.5,0.5) [label={[opacity=\PlaceLabelOpacity]below left:$s_{22}$}] {};

\fill (1.6,0.6) circle (2.5pt);
\fill (1.4,0.4) circle (2.5pt);

\draw[STarc, line width=0.06cm](s21)edge[](t21);
\draw[STarc, line width=0.06cm](s21)edge[](t22);
\draw[STarc](s22)edge[](t21);
\draw[STarc](s22)edge[](t22);

\node[clusterbox, minimum height=2cm, minimum width=3.2cm] (C3) at (6.3,3.6) {};
\node[clusterlabel] at (7.3,5.1) {C3};

\node[transition, fill=pnblue, line width=0.06cm] (t31) at (5,3) [label={[opacity=\TransitionLabelOpacity]left:$t_{31}$}] {};
\node[transition, fill=pnblue] (t32) at (6.5,3) [label={[opacity=\TransitionLabelOpacity]right:$t_{32}$}] {};

\node[place, fill=pnblue, line width=0.06cm] (s31) at (5,4.3) [label={[opacity=\PlaceLabelOpacity, inner sep=0.1mm, label distance=0.2mm]220:$s_{31}$}] {};
\node[place] (s32) at (6.5,4.3) [label={[opacity=\PlaceLabelOpacity, inner sep=0.1mm, label distance=0.2mm]170:$s_{32}$}] {};
\node[place] (s33) at (7.5,4.3) [label={[opacity=\PlaceLabelOpacity]right:$s_{33}$}] {};

\draw[STarc, line width=0.06cm](s31)edge[](t31);
\draw[STarc](s31)edge[](t32);
\draw[STarc](s32)edge[](t31);
\draw[STarc](s32)edge[](t32);
\draw[STarc](s33)edge[](t31);
\draw[STarc](s33)edge[](t32);

\node[clusterbox, minimum height=2cm, minimum width=3.1cm] (C4) at (9.2,0.7) {};
\node[clusterlabel] at (11.3,0.8) {C4};

\node[transition, fill=pnblue, line width=0.06cm] (t41) at (8,0) [label={[opacity=\TransitionLabelOpacity]left:$t_{41}$}] {};
\node[transition, fill=pnblue] (t42) at (9,0) [label={[opacity=\TransitionLabelOpacity]-50:$t_{42}$}] {};
\node[transition, fill=pnblue] (t43) at (10,0) [label={[opacity=\TransitionLabelOpacity]right:$t_{43}$}] {};

\node[place, fill=pnblue, line width=0.06cm] (s41) at (9,1.5) [label={[opacity=\PlaceLabelOpacity]120:$s_{41}$}] {};
\node[place] (s42) at (10,1.5) [label={[opacity=\PlaceLabelOpacity]0:$s_{42}$}] {};

\fill (10,1.5) circle (2.5pt);

\draw[STarc, line width=0.06cm](s41)edge[](t41);
\draw[STarc](s41)edge[](t42);
\draw[STarc](s41)edge[](t43);
\draw[STarc](s42)edge[](t41);
\draw[STarc](s42)edge[](t42);
\draw[STarc](s42)edge[](t43);

\begin{scope}[shift={(-3,4)}]
\node[clusterbox, minimum height=1.6cm, minimum width=2.3cm] (C5) at (12.7,2.5) {};
\node[clusterlabel] at (11.7,3.7) {C5};

\node[transition] (t51) at (12,2) [label={[opacity=\TransitionLabelOpacity]left:$t_{51}$}] {};
\node[transition] (t52) at (13,2) [label={[opacity=\TransitionLabelOpacity]right:$t_{52}$}] {};

\node[place] (s51) at (12,3) [label={[opacity=\PlaceLabelOpacity]left:$s_{51}$}] {};

\fill (12,3) circle (2.5pt);

\draw[STarc](s51)edge[](t51);
\draw[STarc](s51)edge[](t52);
\end{scope}


\draw[TSarc](t12.north east)to[out=20,in=160,looseness=4](s12.north west);
\draw [TSarc](t41.south west) 
 .. controls ++(15:-1.5) and ++(170:-1.5) .. (0,-1)
 .. controls ++(170:1.5) and ++(170:1.5) .. (s12.west);
\draw[TSarc, bend left=40, line width=0.06cm](t21)edge[](s11);

\draw[TSarc, bend left=45, line width=0.06cm](t12)edge[](s21);
\draw[TSarc, bend left=40, line width=0.06cm](t31)edge[](s21);
\draw[TSarc](t42) .. controls +(0,-2) and +(0.5,-2) ..(s22);
\draw[TSarc, bend left=25, line width=0.06cm](t11)to[](s21);

\draw[TSarc, bend left=55, line width=0.06cm](t22)edge[](s31);
\draw[TSarc, bend left=30, line width=0.06cm](t11)edge[](s31);
\draw[TSarc, bend left=45](t11)edge[](s32);
\draw[TSarc, bend left=10](t51)edge[](s33);

\draw[TSarc, line width=0.06cm](t41) .. controls +(-2,1) and +(-4,0.5) .. (s41);
\draw[TSarc](t32)to[out=300,in=120](s42);
\draw[TSarc](t52) .. controls +(1,-2) and +(1,1) .. (s42);

\draw[TSarc](t43) .. controls +(3,-4) and +(4,4) .. (s51);
\draw[TSarc](t51) .. controls +(-2,-2) and +(-2,2) .. (s51);
\draw[TSarc](t52) .. controls +(2,-2) and +(2,1.1) .. (s51);

\end{tikzpicture}

%% file: figures/commoner.tex



\definecolor{pnblue}{RGB}{109, 150, 242}

\newboolean{STarcs}
\newboolean{TSarcs}
\newboolean{ShowPlaceLabels}
\newboolean{ShowTransitionLabels}
\newboolean{ShowPlaces}
\newboolean{ShowTransitions}

\pgfkeys{/mysettings/opacity/.initial=1}

\setboolean{STarcs}{true}
\setboolean{TSarcs}{true}
\setboolean{ShowPlaceLabels}{true}
\setboolean{ShowTransitionLabels}{true}
\setboolean{ShowPlaces}{true}
\setboolean{ShowTransitions}{true}

\ifthenelse{\boolean{STarcs}}{
  \def\STopacity{1}
}{
  \def\STopacity{0}
}
\ifthenelse{\boolean{TSarcs}}{
  \def\TSopacity{1}
}{
  \def\TSopacity{0}
}

\ifthenelse{\boolean{ShowPlaceLabels}}{
  \def\PlaceLabelOpacity{1}
}{
  \def\PlaceLabelOpacity{0}
}
\ifthenelse{\boolean{ShowTransitionLabels}}{
  \def\TransitionLabelOpacity{1}
}{
  \def\TransitionLabelOpacity{0}
}
\ifthenelse{\boolean{ShowPlaces}}{
  \def\PlaceOpacity{1}
}{
  \def\PlaceOpacity{0}
}
\ifthenelse{\boolean{ShowTransitions}}{
  \def\TransitionOpacity{1}
}{
  \def\TransitionOpacity{0}
}

\tikzset{
  STarc/.style={-latex, thick, bend left=0, opacity=\STopacity},
  TSarc/.style={-latex, thick, opacity=\TSopacity},
  place/.style={
  	circle,
  	draw,
  	thick,
  	minimum size=0.5cm,
  	opacity=\PlaceOpacity
  },
  transition/.style={
    draw,
    thick,
    minimum size=0.4cm,
    opacity=\TransitionOpacity
  },
  clusterlabel/.style={
    label=below:#1
  },
  clusterbox/.style={
    rectangle,
    color=blue,
    opacity=0.7,
    fill=yellow,
    rounded corners,
    text width=0.5cm
  }
}


\begin{tikzpicture}[yscale=0.8,xscale=0.85] 

\node at (-6.2,4.5) {$N$};
\draw[ultra thick] (0,0) ellipse (7cm and 6cm);

\node at (-7.5,0) {$R$};
\draw[ultra thick]
  (-7,0)
  arc[
    start angle=170,
    end angle=10,
    x radius=7.1cm,
    y radius=1.5cm
  ];

\node[transition, fill=pnblue] (t11) at (-1.7,2.4) [label={[opacity=\TransitionLabelOpacity]right:$t_{11}$}] {};
\node[transition, line width=0.06cm] (t0) at (-5.8,1.6) [label={[opacity=\TransitionLabelOpacity]above:$t_{0}$}] {};

\node[place] (s1) at (-3.5,2) {};
\node[place] (s2) at (-2,4.3) {};
\node[place] (s3) at (0,4) {};
\node[place] (s16) at (-4.6,3) {};

\node[transition, fill=pnblue] (t12) at (2,1.7) [label={[opacity=\TransitionLabelOpacity]above:$t_{12}$}] {};
\node[place] (s5) at (0.5,2.7) {};
\node[place] (s6) at (5,3) {};
\node[place] (s7) at (3.1,3.8) {};

\node[place] (s13) at (5,1.5) {};

\node at (8.4,-1) {$\pre{T_1} \cap R$};
\draw[thick]
  (-6.6,-2)
  arc[
    start angle=180,
    end angle=0,
    x radius=6.6cm,
    y radius=1.5cm
  ];

\node[place] (s4) at (-1.5,0.2) [label={[opacity=\PlaceLabelOpacity]right:$s_{11}$}] {};
\node[place] (s8) at (1.5,0.2) [label={[opacity=\PlaceLabelOpacity]left:$s_{12}$}] {};
\node[place] (s17) at (0,0.8) {};

\node[transition, fill=pnblue] (t22) at (3.3,0.1) [label={[opacity=\TransitionLabelOpacity]above:$t_{22}$}] {};
\node[transition, fill=pnblue] (t21) at (-3.6,0.1) [label={[opacity=\TransitionLabelOpacity]180:$t_{21}$}] {};

\node at (8,-2.9) {$\pre{T_2} \cap R$};
\draw[thick]
  (-5.7,-3.5)
  arc[
    start angle=160,
    end angle=20,
    x radius=6.05cm,
    y radius=1.5cm
  ];

\node[place] (s9) at (-4.8,-1.7) [label={[opacity=\PlaceLabelOpacity]left:$s_{21}$}] {};
\node[place] (s10) at (2.5,-1.7) [label={[opacity=\PlaceLabelOpacity]below:$s_{22}$}] {};
\node[place] (s14) at (-1,-1.7) {};

\node[transition, fill=pnblue] (t32) at (4.3,-2) [label={[opacity=\TransitionLabelOpacity]above:$t_{32}$}] {};
\node[transition, fill=pnblue] (t31) at (-3,-1.7) [label={[opacity=\TransitionLabelOpacity]above:$t_{31}$}] {};

\node at (6.2,-4.5) {$\pre{T_3} \cap R$};

\node[place] (s31) at (-3.6,-3.5) {};
\node[place] (s12) at (4.5,-3.5) [label={[opacity=\PlaceLabelOpacity]left:$s_{32}$}] {};
\node[place] (s15) at (-2.1,-3.1) [label={[opacity=\PlaceLabelOpacity]right:$s_{31}$}] {};

\node at (3,-4.7) {$Q$};
\draw[thick]
  (-3.9,-5)
  arc[
    start angle=170,
    end angle=10,
    x radius=3.95cm,
    y radius=2cm
  ];

\node[place] (sq) at (-2,-4.5) {};


\draw[STarc](s1)edge[](t11);
\draw[TSarc](t11)edge[](s3);
\draw[TSarc, bend right=50](t11)edge[](s16);
\draw[TSarc, bend right=10](t11)edge[](s2);
\draw[TSarc](t12)edge[](s6);
\draw[TSarc, bend right=30](t12)edge[](s7);
\draw[STarc](s5)edge[](t12);

\draw[STarc](s4)edge[](t11);
\draw[STarc](s8)edge[](t12);
\draw[STarc, bend left=20](s13)edge[](t22);

\draw[TSarc](t21)edge[](s4);
\draw[TSarc](t21)edge[](s1);
\draw[TSarc](t22)edge[](s8);

\draw[STarc](s9)edge[](t21);
\draw[STarc, bend left=50](s6)edge[](t32);
\draw[STarc](s9)edge[](t0);
\draw[STarc](s10)edge[](t22);
\draw[STarc](s17)edge[](t12);
\draw[STarc](s16)edge[](t0);
\draw[STarc, bend right=20](s16)edge[](t21);

\draw[TSarc](t31)edge[](s9);
\draw[TSarc](t32)edge[](s10);
\draw[TSarc](t31)edge[](s14);

\draw[STarc](s12)edge[](t32);
\draw[STarc](s31)edge[](t31);
\draw[STarc](s15)edge[](t31);
\draw[STarc, bend right=10](s14)edge[](t22);

\draw[TSarc, bend right=60](t0)edge[](sq);

\end{tikzpicture}